\documentclass[prb,superscriptaddress]{revtex4}            %% REVTeX 4.0
\usepackage{amsmath}
\usepackage{amsfonts}
\usepackage[dvips]{graphicx}
\usepackage{epsfig}
\newcommand{\eref}[1]{(\ref{#1})}
\newcommand{\sref}[2]{(\ref{#1}-\ref{#2})}
\newcommand{\tref}[2]{(\ref{#1},\ref{#2})}
\newcommand{\Eref}[1]{Eq.~(\ref{#1})}
\newcommand{\Mref}[2]{Eqs.~(\ref{#1}-\ref{#2})}
\newcommand{\Tref}[2]{Eqs.~(\ref{#1},\ref{#2})}
\newcommand{\Fref}[1]{Fig.~\ref{#1}}

\newcommand{\Aref}[1]{Appendix ~\ref{#1}}
\newcommand{\rmi}{\mathrm i}
\newcommand{\rme}{\exp}
\newcommand{\e}{\rme}
\newcommand{\rmd}{\mathrm d}
\renewcommand{\vec}{\boldsymbol}
\newcommand{\citeo}[1]{Ref.~\onlinecite{#1}}
\newcommand{\degre}{^{\mathrm o}}
\newcommand{\ep}{\epsilon}
\newcommand{\x}{\vec{x}}
\newcommand{\E}{\vec{E}}
\newcommand{\p}[1]{\vec{p}_{#1}}
\newcommand{\q}[1]{\vec{q}_{#1}}
\newcommand{\op}[1]{\overline{\vec{#1}}}
\newcommand{\hx}{h(\vec{x})}
\newcommand{\rx}{\vec{r}_{\x}}
\newcommand{\kp}[2]{\vec{k}_{\p{#1}}^{#2}}
\newcommand{\ku}[2]{\vec{k}_{\vec{u}{#1}}^{#2}}
\newcommand{\hvec}[1]{\hat{\vec{#1}}}
\newcommand{\ehp}[1]{\hvec{e}_H(\p{#1})}
\newcommand{\evp}[2]{\hvec{e}_V^{#1}(\p{#2})}
\newcommand{\ehu}[1]{\hvec{e}_H(\vec{u}{#1})}
\newcommand{\evu}[2]{\hvec{e}_V^{#1}(\vec{u}{#2})}
\newcommand{\nx}{\vec{n}(\x)}
\newcommand{\ex}{\hvec{e}_x}
\newcommand{\ey}{\hvec{e}_y}
\newcommand{\ez}{\hvec{e}_z}
\newcommand{\lap}{\nabla^2}
\newcommand{\intp}[1]{\int \frac{\rmd^2 \vec{p}_{#1}}{(2\pi)^2}}
\newcommand{\iintp}[2]{\int \!\!\! \int \frac{\rmd^2 \vec{p}_{#1}}{(2\pi)^2} 
\frac{\rmd^2 \vec{p}_{#2}}{(2\pi)^2}}
\newcommand{\iiintp}[3]{\int \!\!\! \int \!\!\! \int \frac{\rmd^2 \vec{p}_{#1}}
{(2\pi)^2} 
\frac{\rmd^2 \vec{p}_{#2}}{(2\pi)^2} \frac{\rmd^2 \vec{p}_{#3}}{(2\pi)^2}}
\newcommand{\iintxp}[1]{\int \!\!\! \int \rmd^2 \vec{x} \frac{\rmd^2 
\vec{p}_{#1}}{(2\pi)^2}}
\newcommand{\ga}{\gamma}
\newcommand{\alp}[2]{\alpha_{#1}(\vec{p}_{#2})}
\newcommand{\alu}[2]{\alpha_{#1}(\vec{u}_{#2})}
\newcommand{\Rp}[2]{\op{R}(\p{#1}|\p{#2})}
\newcommand{\no}{\nonumber}
\newenvironment{vecteur}{\left( \begin{array}{c}}{\end{array} \right)}
\newenvironment{matrice}{\begin{pmatrix}}{\end{pmatrix}}
\begin{document}
\title{A new application of reduced  Rayleigh equations to 
electromagnetic wave scattering by
two-dimensional randomly rough surfaces}
\author{Antoine Soubret}
\email{soubret@cpt.univ-mrs.fr}
\affiliation{Thomson-CSF Optronique, 
BP 55, 78233 Guyancourt Cedex, France.}
\affiliation{Centre de Physique Th\'eorique, CNRS-Luminy Case
907, 13288 Marseille Cedex 9, France.}
\author{G\'erard Berginc}
\affiliation{Thomson-CSF Optronique, 
BP 55, 78233 Guyancourt Cedex, France.}
\author{Claude  Bourrely}
\affiliation{Centre de Physique Th\'eorique, CNRS-Luminy Case
907, 13288 Marseille Cedex 9, France.}
\date{September 2000}
%%%%%%%
\begin{abstract}
The small perturbations method has been extensively used for waves
scattering by rough surfaces. The standard method developped
by Rice is difficult to apply when we consider
second and third order of scattered fields as a function of the surface
height. Calculations can be greatly simplified with the use of reduced Rayleigh 
equations, because one of the unknown fields can be eliminated. 
We derive a new set of four reduced equations for the scattering amplitudes, 
which are applied to the cases of a rough conducting surface, 
and to a slab where one of the boundary is a rough surface. As in the
one-dimensional case, numerical simulations show the appearance of 
enhanced backscattering for these structures.
\end{abstract}

\pacs{PACS number(s): 42.25.Fx, 42.25.Ja, 42.25.Gy, 78.66.Bz, 78.20.e, 73.20.Mf} 

\maketitle    

%%%%%%%%%%%%%%%%%%%%%%%%%%%%%%%%%%%%%%%%%%%%%%%%%%%%%%%%%%%%%%%%%%%%%%%%%%%%%%%
\section{Introduction}
%%%%%%%%%%%%%%%%%%%%%%%%%%%%%%%%%%%%%%%%%%%%%%%%%%%%%%%%%%%%%%%%%%%%%%%%%%%%%%%
The scattering of electromagnetic waves from a rough surface has been
studied in different domains such as radio-physics, geophysical remote
sensing, ocean acoustics, and surface optics.~\cite{Bass,Ishi,
Ulaby,Kong,Desanto,Ogilvy,Kravtsov,Voro,Fung,NietoL,Mara} 
One of the earlier theory used, is the small perturbations method (SPM)
originally developped by Rice,~\cite{Rice} this theory still remains 
of interest~\cite{Mara,Johnson,Fuks} because perturbative terms 
of order higher than one can
produce enhanced backscattering, or improve predictions accuracy in an
emission model. Although, the Rice method can be used in principle
to determine all orders in the perturbative development, 
very few works use terms of order higher
than one for a two-dimensional surface due to the calculations
complexity. The second order has been written in a compact form by
Voronovich~\cite{VoroSSA} in his work on small-slope approximation, 
and only recently the third order has been presented.~\cite{Johnson}
However, there exists a different way to obtain (SPM) which dates back 
from the work of Brown {\it et al.}~\cite{Celli}
Using both the Rayleigh hypothesis and 
the extinction theorem, they have obtained an integral equation, called
the reduced Rayleigh equation, which only involves the incident and
the scattered field alone. In their method the field transmitted 
through the surface has been eliminated 
in such a way that the scattered field becomes a
function of the incident field only. This reduced equation has been
extensively used by Maradudin {\it et al.} to study localization 
effects by a conducting surface,~\cite{Mara1,Mara2,Mara3} coherent
effects in reflection factor,~\cite{Baylard} scattering by
one-dimensional~\cite{Mendez} and two-dimensional conducting
surface.~\cite{McGurn} It has to be noticed that the third order
perturbation term was already explicited in the work of \citeo{McGurn}.

In recent years, similar studies have been done in the case of thin
films bounded by a rough surface,~\cite{Sanchez,Mara} but only in the
one-dimensional case.\footnote{However, we have to mention the work
Ref. 27 where they consider a two-dimensional film. Their method
has in principle the same domain of validity as the (SPM), 
the difference comes from their use of a
sophisticated development called Wiener-Hermitte expansion to write down
the scattered fields.} In order to calculate the two-dimensional case
it becomes necessary to derive an extension of the reduced Rayleigh
equation for this system. In the present paper, we first study a
surface where a down-going and up-going  fields exist both on the upper
and the bottom side of the film. 
We show the existence of four equations, that we
also call reduced Rayleigh equation, they have the property that
one of the down-going or up-going fields has been eliminated. 
With these equations, we rediscover the equation obtained by Brown 
{\it et al.}~\cite{Celli} for one rough
surface, and derive the corresponding ones for a slab where one of the
boundary is a rough surface.
Next, a perturbative development up to third order is obtained in a
compact matrix form for these two systems. This third order term is
mandatory if we need the expression of the cross section up to fourth order
approximation. 
As in the one-dimensional case, the results of the incoherent 
cross section show a well defined peak in the retroreflection direction. 

In the case of small-roughness metallic surfaces this peak was originaly 
explained by the infinite perturbation theory,~\cite{Mara1,Mara2,Mara3} 
and further developments~\cite{Mendez}
have shown that the major contribution for the enhanced
backscattering peak comes from the second order term in the field
perturbation. However, in the one dimensional case 
the enhanced backscattering
for a rough surface, appears only for a $(TM)$ incident wave due to the
fact that plasmons polaritons only exist for this polarization. In the 
two dimensional case, due to the existence of cross-polarization, 
we will show that an incident $(TE)$ wave 
can excite a $(TM)$ plasmon mode which can transform into a
$(TE)$ or $(TM)$ volume electromagnetic wave. Thus, the enhanced
backscattering is present independently of the polarization of the incident
and scattered waves. For a rough dielectric film bounded by a
conducting plane the enhanced backscattering is present for both $(TE)$
and $(TM)$ incident waves, even in the one-dimensional case because guided 
waves exist for these two polarizations. The qualitative effect of
the two-dimensional surface is particularly sensitive when we study
thin film, for instance,
in one-dimensional case, satellite peaks~\cite{Sanchez,Mara} 
appear on each side of the enhanced backscattering peaks, 
however, in the two-dimensional case, the
coupling between $(TE)$ and $(TM)$ modes attenuates drastically these peaks.

The paper is organized as follows. In Sec.~\ref{RedRay}, we 
derive the four new reduced Rayleigh equations. In Sec.~\ref{Rmatrix}, we
introduce the diffusion matrix, and in Sec.~\ref{Pertu}, we
determine the perturbative development up to third order term in the
surface height, for a rough surface alone, and for slab with a 
rough surface located at one of the boundaries. 
In Sec.~\ref{Muller}, we introduce the Mueller
matrix and the definition of the statistical parameters for the 
rough surface.
We then obtain the bistatic matrix in terms of a perturbative development.
Numerical examples which show the enhanced backscattering are
presented in Sec.~\ref{Numer}. Conclusions drawn from the results of 
our calculations are discussed in Sec.~\ref{Concl}.

%%%%%%%%%%%%%%%%%%%%%%%%%%%%%%%%%%%%%%%%%%%%%%%%%%%%%%%%%%%%%%%%%%%%%%%%%%%%%%%
\section{Derivation of the reduced Rayleigh equation}
\begin{figure}[ht]
\input{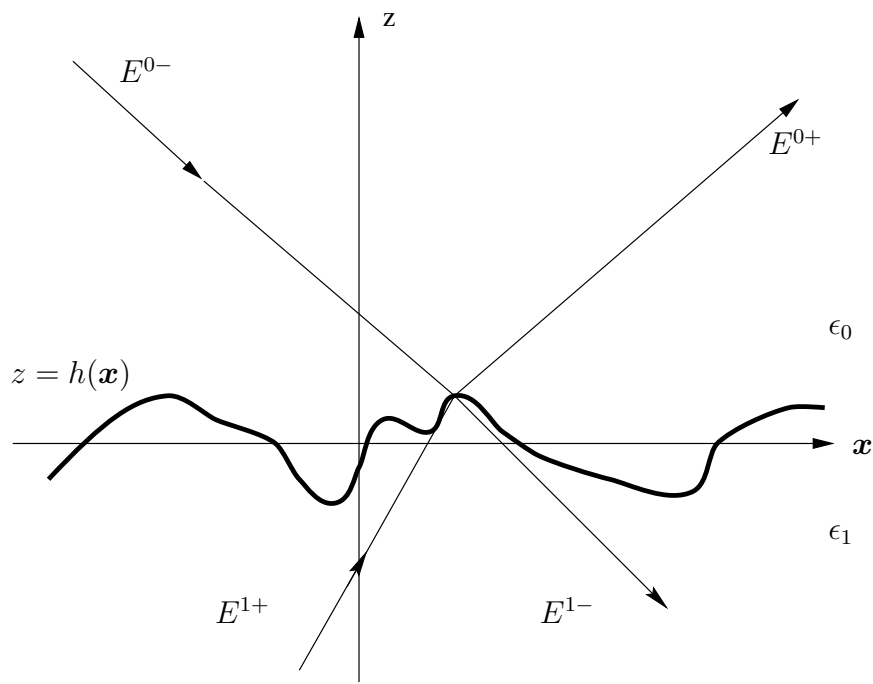}
\caption{A rough surface with an incident wave coming from both 
sides of medium 0 and 1.}
\label{surface}
\end{figure}

%%%%%%%%%%%%%%%%%%%%%%%%%%%%%%%%%%%%%%%%%%%%%%%%%%%%%%%%%%%%%%%%%%%%%%%%%%%%%%%
\label{RedRay}
The reduced Rayleigh equation was obtained for a two-dimensional
surface by Brown {\it et al.}~\cite{Celli} using the extinction theorem
and the Rayleigh hypothesis, it allows to calculate the
scattered field from the rough surface. Now, if we want to compute
the transmitted field  
from the rough surface, one has to introduce an other reduced Rayleigh equation
derived by Greffet.~\cite{Greffet} However, these two equations were
established in the case where there is no up-going  field
inside the medium, thus they cannot be used to obtain the field scattered 
by a slab with a rough surface in its upper side. In fact, to 
generalize these equations to a slab, we have to consider all the fields
shown in \Fref{surface}. We will prove that there
exists four reduced Rayleigh equations, which involve only three of the
participating fields $\E_0^{-},\E_0^{+},\E_1^{-},\E_1^{+}$.

We consider that each electromagnetic waves propagates with a
frequency $\omega$,
and in the following the factor $exp(-i\omega t)$ will be omitted.
We choose to work with a Cartesian coordinate system
$\vec{r}=(\vec{x},z)=(x,y,z)$,
where the z axis directed upward, and we consider a boundary of the 
form $z=h(\vec{x})$. Moreover, we suppose that there exists a length
$L$ for which $\,h(x,y)=0$, if $|x|>L/2$ or $|y|>L/2$, $L$ may be
arbitrary large but finite.

%%%%%%%%%%%%%%%%%%%%%%%%%%%%%%%%%%%%%%%%%%%%%%%%%%%%%%%%%%%%%%%%%%%%%%%%%%%%%%%
\subsection{Propagation equations and boundary conditions}
\begin{figure}[ht]
\input{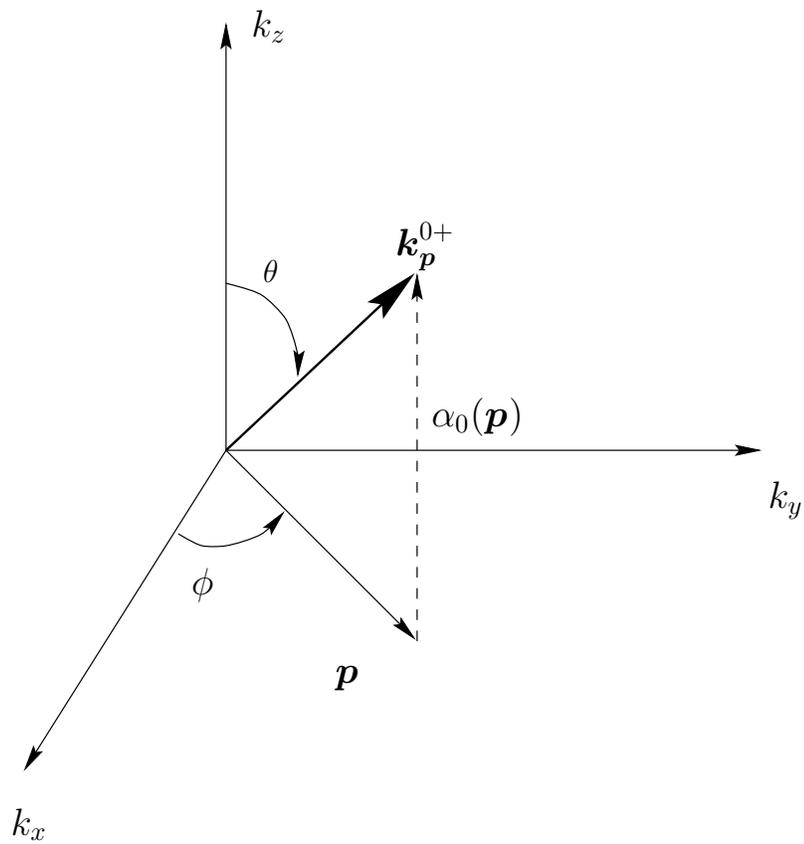}
\caption{Decomposition of the wave vector $\kp{}{0+}$.}
\label{Vectonde}
\end{figure}

%%%%%%%%%%%%%%%%%%%%%%%%%%%%%%%%%%%%%%%%%%%%%%%%%%%%%%%%%%%%%%%%%%%%%%%%%%%%%%%
The electric field $\vec{E}$ satisfies the Helmohltz equation in the two media:
\begin{eqnarray}
&&(\lap+\ep_0 K_0^2)\vec{E}^0(\vec{r})=0 \quad \mbox{for }
z>h(\vec{x})\, , \\
&&  (\lap+\ep_1 K_0^2)\vec{E}^1(\vec{r})=0 \quad \mbox{for }
z<h(\vec{x})\, , 
\end{eqnarray} 
where $K_0=\omega/c$.
Since the system is homogeneous in the $\vec{x}=(x,y)$ directions, we can
represent the electric field by its Fourier transform. Thus, using the
Helmohltz equation, we deduce the following expression for the 
electric field~\cite{Voro,NietoL} in the medium 0 :
\begin{equation}
\vec{E}^{0}(\vec{r})=\intp{}\vec{E}^{0-}(\p{})\e(\rmi\kp{}{0-}\cdot\vec{r} ) 
+\intp{}\vec{E}^{0+}(\p{})\e(\rmi\kp{}{0+}\cdot\vec{r} )\, , \label{DefE0}
\end{equation}
where (see \Fref{Vectonde})
\begin{align}
\alp{0}{}\equiv &(\ep_0K_0^2-\p{}^2)^{\frac{1}{2}} \, , \label{alpha0}\\
\kp{}{0\pm}\equiv & \vec p\pm\alpha_0(\vec p)\ez \, .
\end{align}
In fact, when writing such a definition, 
we made implicitely the assumption that the Rayleigh
hypothesis is correct. This representation is only valid when
$z>max[h(\x)]$ and in that case $\vec{E}^{0-}(\p{})$ represents
the incident wave amplitude. In order to be correct we need to add an 
explicit dependence in the $z$ coordinate like (see \citeo{Brown}):
\begin{equation}
\vec{E}^{0-}=\vec{E}^{0-}(\p{},z)  \qquad \vec{E}^{0+}=\vec{E}^{0+}(\p{},z)
\end{equation}
However, explicit calculations in the case of infinite
conducting surfaces,~\cite{Nieto} and for a dielectric medium~\cite{Kong} 
without this hypothesis, have shown that the perturbative developments are
identical. The validity of this hypothesis is by no doubt a matter
of convergence domain as discussed by Voronovich.~\cite{Voro,VoroSSA}

In the medium 1, we have a similar expression :
\begin{equation}
\vec{E}^{1}(\vec{r})=\intp{}
\vec{E}^{1-}(\vec{p})\e(\rmi\kp{}{1-}\cdot\vec{r} )
+\intp{}\vec{E}^{1+}(\p{})\e(\rmi\kp{}{1+}\cdot\vec{r})\, ,
\label{DefE1}
\end{equation}
where
\begin{align}
\alp{1}{}\equiv & (\ep_1K_0^2-\p{}^2)^{\frac{1}{2}} \, ,\label{alpha1}\\
\vec{k}_{\p{}}^{1\pm}\equiv & \vec p\pm\alpha_1(\vec p)\ez \, .
\end{align}
We decompose the vectors $\vec{E}(\p{})$  on a two-dimensional basis due
to the fact that $\nabla.\vec{E}(\vec{r})=0$, which gives the
conditions :
\begin{equation}
\vec{k}_{\p{}}^{0\pm}.\vec{E}^{0\pm}(\p{})=0,  \qquad
\vec{k}_{\p{}}^{1\pm}.\vec{E}^{1\pm}(\p{})=0\, .
\end{equation}
Then, we define the horizontal polarization vectors $H$ for $(TE)$ and
$V$ for $(TM)$ in medium 0 by :
\begin{align}\ehp{}& \equiv \frac{\ez \times \vec k_{\p{}}^{0\pm}}{||\ez 
\times \vec k_{\p{}}^{0\pm}||}=\ez \times
\hvec{p} \, ,
\label{TE0}\\
\evp{0\pm}{} & \equiv   \frac{\ehp{} \times \vec k_{\p{}}^{0\pm}}{||\ehp{0} 
\times \vec k_{\p{}}^{0\pm}||}=\pm
\frac{\alpha_0(\vec p)}{\sqrt{\ep_0}K_0}\hvec{p}-\frac{||\vec p||}
{\sqrt{\ep_0}K_0}\ez \, ,
\label{TM0}
\end{align}
with similar expressions for the medium 1:
\begin{align}\ehp{}& \equiv \frac{\ez \times \vec k_{\p{}}^{1\pm}}{||\ez 
\times \vec k_{\p{}}^{1\pm}||}=\ez \times
\hvec{p} \, ,
\label{TE1}\\
\evp{1\pm}{} & \equiv   \frac{\ehp{} \times \vec k_{\p{}}^{1\pm}}{||\ehp{0} 
\times \vec k_{\p{}}^{1\pm}||}=\pm
\frac{\alpha_1(\vec p)}{\sqrt{\ep_1}K_0}\hvec{p}-
\frac{||\vec p||}{\sqrt{\ep_1}K_0}\ez \, .
\label{TM1}
\end{align}
So, we decompose the waves in medium 0 on the basis
$[\p{}]^{0-}\equiv (\evp{0-}{},\ehp{})$, and $[\p{}]^{0+}\equiv 
(\evp{0+}{},\ehp{})$:
\begin{equation}
\vec{E}^{0-}(\p{})=\begin{vecteur} E^{0-}_V(\p{}) \\
E_H^{0-}(\p{})\end{vecteur}_{[p]^{0-}},\qquad
\vec{E}^{0+}(\p{})=\begin{vecteur} E^{0+}_V(\p{}) \\
E_H^{0+}(\p{})\end{vecteur}_{[p]^{0+}} \, ,
\end{equation}
and for medium 1 on the basis $[\p{}]^{1-}\equiv (\evp{1-}{},\ehp{})$
and $[\p{}]^{1+}\equiv (\evp{1+}{},\ehp{})$:
\begin{equation}
\vec{E}^{1-}(\p{})=\begin{vecteur} E^{1-}_V(\p{}) \\
E_H^{1-}(\p{})\end{vecteur}_{[p]^{1-}} ,\qquad 
\vec{E}^{1+}(\p{})=\begin{vecteur} E^{1+}_V(\p{}) \\
E_H^{1+}(\p{})\end{vecteur}_{[p]^{1+}} \, .
\end{equation}
The electric $\vec{E}(\vec{x},z)$ and magnetic fields
$\vec{B}(\vec{x},z)=\frac{1}{i\omega}\nabla\times\vec{E}(\vec{x},z)$,
satisfy the following boundary conditions :
\begin{align}
\vec n(\vec{x})\times\left[\vec E^{0}(\vec x,h(\vec{x}))-\vec E^{1}(\vec
x,h(\x))\right]&=0 \label{CCL1} \, ,\\ \vec n(\x)\cdot \left[\ep_0\vec 
E^0(\vec x,h(\vec{x}))-\ep_1\vec E^1(\vec
x,h(\vec x))\right]&=0 \label{CCL2} \, ,\\
\vec n(\x)\times\left[\vec B^0(\vec x,h(\vec{x}))-\vec
B^{1}(\x,h(\vec{x}))\right]&=0 \label{CCL3} \, ,\\
\vec n(\x)\equiv \ez-\nabla h(\x)\ . & \nonumber
\end{align}
Let us introduce the fields Fourier transform, \Eref{DefE0} and
\Eref{DefE1}, in the boundary conditions \Mref{CCL1}{CCL3}, they give :
%\Eref{CCL1},\Eref{CCL2} and \Eref{CCL3} which gives :
\begin{align}
\sum_{a=\pm}\intp{} \vec{n}(\x)\times
\vec{E}^{0a}(\p{})\e(i\kp{}{0a}\cdot\rx)&=
\sum_{a=\pm}\intp{} \vec{n}(\x)\times
\vec{E}^{1a}(\p{})\e(i\kp{}{1a}\cdot\rx) \, ,\label{CL1}\\
\frac{\ep_0}{\ep_1}\sum_{a=\pm}\intp{} \vec{n}(\x)\cdot
\vec{E}^{0a}(\p{})\e(i\kp{}{0a}\cdot\rx)&=
\sum_{a=\pm}\intp{} \vec{n}(\x)\cdot
\vec{E}^{1a}(\p{})\e(i\kp{}{1a}\cdot\rx) \, ,\label{CL2}\\
\sum_{a=\pm}\intp{} \vec{n}(\x)\times[\vec{k}_{\p{}}^{0a}\times
\vec{E}^{0a}(\p{})]\e(i\kp{}{0a}\cdot\rx)&=
\sum_{a=\pm}\intp{} \vec{n}(\x)\times[\vec{k}_{\p{}}^{1a}\times
\vec{E}^{1a}(\p{})]\e(i\kp{}{1a}\cdot\rx) \, ,\label{CL3}\\
\rx=\x+\hx\ez \, , \qquad
\kp{}{0a}\equiv \vec{p}+a\alp{0}{}\ez, & \qquad  \kp{}{1a}\equiv
\vec{p}+a\alp{1}{}\ez \, ,
\end{align}
where the summation includes the two possible signs: $a=\pm$, linked to the
propagation directions.
We will also use the condition
$\nabla\cdot\vec{E}^{0}(\x,z)=\nabla\cdot\vec{E}^{1}(\x,z)$, 
which gives the relation
\begin{equation}
\sum_{a=\pm}\intp{} \vec{k}_{\p{}}^{0a}\cdot
\vec{E}^{0a}(\p{})\e(i\kp{}{0a}\cdot\rx)=
\sum_{a=\pm}\intp{} \vec{k}_{\p{}}^{1a}\cdot
\vec{E}^{1a}(\p{})\e(i\kp{}{1a}\cdot\rx) \label{CL4} \, .
\end{equation}

%%%%%%%%%%%%%%%%%%%%%%%%%%%%%%%%%%%%%%%%%%%%%%%%%%%%%%%%%%%%%%%%%%%%%%%%%%%%%%%
\subsection{Fields elimination}
%%%%%%%%%%%%%%%%%%%%%%%%%%%%%%%%%%%%%%%%%%%%%%%%%%%%%%%%%%%%%%%%%%%%%%%%%%%%%%%
The equations \sref{CL1}{CL3} and \eref{CL4}, are all
linear in the fields $\E^{0-}$, $\E^{0+}$, $\E^{1-}$, $\E^{1+}$. 
In order to eliminate  $\E^{1-}$ or $\E^{1+}$ in the equations
\sref{CL1}{CL3} and \eref{CL4}, we will take the
following linear combination of their left and right members:
\begin{equation}
\int \rmd^2 \x \, [\ku{}{1b}\times
(\Eref{CL1})+(\Eref{CL3})-\ku{}{1b}(\Eref{CL2})-\vec{n}(\x)(\Eref{CL4})]
\e(-i\ku{}{1b}\cdot\rx)\, ,
\label{LL}
\end{equation}
with
$\ku{}{1b}\equiv \vec{u}+b\alu{1}{}\ez$,
and where $b=\pm$, has to be fixed according to the choice of the field we want
to eliminate.
With the vectorial identity,
$\vec{a}\times(\vec{b}\times\vec{c})=\vec{b}(\vec{a}\cdot\vec{c})-\vec{c}
(\vec{a}\cdot\vec{b})$, 
the \underline{right member} of \Eref{LL} can be written :
\begin{align}
\sum_{a=\pm}\iintxp{}&\left[-(\ku{}{1b}+\kp{}{1a})\cdot \nx
\, \E^{1a}(\p{})+(\ku{}{1b}-\kp{}{1a})\cdot\E^{1a}(\p{})
\,\nx \right. \nonumber \\ &\left. -\nx\cdot\E^{1a}(\p{})\,
(\ku{}{1b}-\kp{}{1a})\right] \e(-i(\ku{}{1b}-\kp{}{1a})\cdot\rx) \, .
\label{TermeR}
\end{align}
We have now to discuss the different cases depending on the relative 
sign between $a$ and $b$:

1)~If $\underline{a=-b}$, we can use an integration by parts (see
\Aref{IP}) to evaluate $\nx\equiv\ez-\nabla\hx$.
Then we can make the replacement :
\begin{equation}  
\nx=\ez-\nabla \hx \longleftrightarrow
\nx=\ez+\frac{(\vec{u}-\p{})}{(b\alu{1}{}-a\alp{1}{})}\, .
\label{vnx}
\end{equation}
It has to be noticed that the denominator $(b\alu{1}{}-a\alp{1}{})$
does not present any singularity because $a=-b$.
For the first term in the integral \eref{TermeR} we obtain:
\begin{align}
-(\ku{}{1b}+\kp{}{1a})\cdot \nx
\, \E^{1a}(\p{})&=\frac{-b
\E^{1a}(\p{})}{(\alu{1}{}+\alp{1}{})}[\vec{u}^2-\vec{p}^2+\alu{1}{}^2-
\alp{1}{}^2]\, , \no \\
&=0 \, .
\end{align}
The last equality can be easily checked using \Eref{alpha1}.
For the sum of the second and third terms of \Eref{TermeR}, we have
also:
\begin{equation}
(\ku{}{1b}-\kp{}{1a})\cdot\E^{1a}(\p{})
\,\nx  -\nx\cdot\E^{1a}(\p{})\,(\ku{}{1b}-\kp{}{1a})=0 \, ,
\end{equation}
due to the fact that:
\begin{equation}
\nx=\frac{\ku{}{1b}-\kp{}{1a}}{b\alu{1}{}-a\alp{1}{}} \, .
\end{equation}

2)~If $\underline{a=b}$, we can use again the integration by parts only if 
$\alu{1}{}\neq \alp{1}{}$. Then we have to consider three cases:

2-a)~$\vec{u}\neq\vec{p}$ and $\vec{u}\neq-\vec{p}$, as in the previous
case by using an integration by parts we show that \Eref{TermeR} is zero.

2-b)~$\vec{u}=\vec{p}$, then $\ku{}{1b}=\kp{}{1a}$:
\begin{align}
-\int \rmd^2 \x (\ku{}{1b}+\kp{}{1a})\cdot \nx \, \E^{1a}(\p{})\,
\e(-i(\ku{}{1b}-\kp{}{1a})\cdot\rx)&=-\int  \rmd^2 \x
2\ku{}{1b}\cdot \nx \, \E^{1a}(\p{}) \nonumber \\
&=-2b\alu{1}{}\int\rmd \x \no \\
&=-2b\alu{1}{}\, L^2 \, ,
\end{align}
because $\int \rmd^2 \x \nabla \hx=0$,
and 
\begin{equation}
(\ku{}{1b}-\kp{}{1a})\cdot\E^{1a}(\p{})
\,\nx  -\nx\cdot\E^{1a}(\p{})\,(\ku{}{1b}-\kp{}{1a})=0 \, .
\end{equation}

2-c)~$\vec{u}=-\vec{p}\neq 0$, then $\ku{}{1b}-\kp{}{1a}=2\vec{u}\, ,$
\begin{align}
\int \rmd^2 \x \, \nx \,
&\e(-i(\ku{}{1b}-\kp{}{1a})\cdot\rx)=\int \rmd^2 \x \,(\ez-\nabla\hx)\, 
\e(-2i\vec{u}\cdot\x) \nonumber \\
&=\ez \int \rmd^2 \x \e(-2i\vec{u}\cdot\x) \x -\ex \int \rmd
y\left[\e(-2i\vec{u}\cdot\x)h(x,y)\right]_{x=-L/2}^{x=L/2} \no \\ &-\ey 
\int \rmd x\left[\e(-2i\vec{u}\cdot\x)h(x,y)\right]_{y=-L/2}^{y=L/2}
\nonumber \\
&=\ez \delta(\vec{u})\quad \mbox{when} \quad L\rightarrow+\infty \no \\
&=0 \qquad \mbox{since} \quad \vec{u}\neq 0 \, .
\end{align}
This result implies that the expression \eref{TermeR} is also zero 
in that case.

We can summarize all the above results in the form :
\begin{align}
-\int \rmd^2 \x (\ku{}{1b}+\kp{}{1a})\cdot \nx \, \E^{1a}(\p{})\,
\e(-i(\ku{}{1b}-\kp{}{1a})\cdot\rx)&=-2b\alu{1}{}\,\delta_{a,b}\delta_{\vec{u},
\p{}}L^2 \no \\
&=-2b\alu{1}{}\,\delta_{a,b}\,(2\pi)^2\,\delta(\vec{u}-\p{})
\E^{1b}(\vec{u}) \label{Terme1s} \\
& \mbox{when} \quad L\rightarrow+\infty \, , \no
%\quad \mbox{when} \quad L\rightarrow+\infty \, ,
%\label{Terme1s}
\end{align}
where $\delta_{\vec{u},\p{}}$ is the kronecker symbol, and
$\delta(\vec{u}-\p{})=(2\pi)^2/L^2\,\delta_{\vec{u},\p{}}$ the Dirac function.

After an integration on $\p{}$ and a summation on $a$, we obtain for
the expression \eref{TermeR}:
\begin{equation}
-2b\alu{1}{}\E^{1b}(\vec{u})\, .
\label{part35}
\end{equation}
We see that we can eliminate the field $\E^{1-}(\vec{u})$ or
$\E^{1+}(\vec{u})$ depending on the choice made for $b=\pm$.

Now, if we consider the \underline{left member} of \Eref{LL}, we have:
\begin{align}
\sum_{a=\pm}\iintxp{}&\left[-(\ku{}{1b}+\kp{}{0a})\cdot \nx
\, \E^{0a}(\p{})+(\ku{}{1b}-\kp{}{0a})\cdot\E^{0a}(\p{})
\,\nx \right. \nonumber \\ &\left. -\nx\cdot\E^{0a}(\p{})\,
(\frac{\ep_0}{\ep_1}\ku{}{1b}-\kp{}{0a})\right]
\e(-i(\ku{}{1b}-\kp{}{0a})\cdot\rx)\, .
\label{TermL}
\end{align}
Using an integration by parts, we replace $\nx$ by \eref{vnx}:
\begin{align}  
\nx& \longleftrightarrow
\ez+\frac{(\vec{u}-\p{})}{(b\alu{1}{}-a\alp{0}{})}=\frac{\ku{}{1b}-
\kp{}{0a}}{(b\alu{1}{}-a\alp{0}{})} \,.
\end{align}
In this case there is no need to discuss the relative sign between 
$a$ and $b$ because $b\alu{1}{}-a\alp{0}{}\neq 0$, due to the fact
that $\ep_0\neq\ep_1$.
We then obtain :
\begin{align}
-(\ku{}{1b}+\kp{}{1a})\cdot \nx \,
\E^{0a}(\p{})&=-\frac{\vec{u}^2-\p{}^2+\alu{1}{}^2-\alp{0}{}^2}{b\alu{1}{}-
a\alp{0}{}}\,\E^{0a}(\p{})\, , \no \\
&=-\frac{(\ep_1-\ep_0)K_0^2}{b\alu{1}{}-a\alp{0}{}}\,\E^{0a}(\p{})\, ,
\label{part38}
\end{align}
where we have used the definitions \eref{alpha0} and \eref{alpha1}.
The remaining terms of \Eref{TermL} give :
\begin{align}
&(\ku{}{1b}-\kp{}{0a})\cdot\E^{0a}(\p{})
\,\nx-\nx\cdot\E^{0a}(\p{})\,(\frac{\ep_0}{\ep_1}\ku{}{1b}-\kp{}{0a}) \no \\
&=(\ku{}{1b}-\kp{}{0a})\cdot\E^{0a}(\p{})\,\nx 
-\nx\cdot\E^{0a}(\p{})\,(\ku{}{1b}-\kp{}{0a})+\nx\cdot\E^{0a}(\p{})\,
(\ku{}{1b}-\frac{\ep_0}{\ep_1}\ku{}{1b}) 
\no \\
&=\frac{\ku{}{1b}-\kp{}{0a}}{b\alu{1}{}-a\alp{0}{}}\cdot\E^{0a}(\p{})\, 
\frac{(\ep_1-\ep_0)}{\ep_1}\,\ku{}{1b} \, .
\label{part39}
\end{align}
Introducing the following notation :
\begin{equation}
I(\alpha|\p{})\equiv\int \rmd^2\x\, \e(-\rmi
\p{}\cdot\x-i\alpha\,\hx)\, ,
\label{ialp}
\end{equation}
and taking into account the expressions \eref{part35}, 
\sref{part38}{part39},
we express the resulting linear combination \eref{LL} in the form:
\begin{align}
\sum_{a=\pm}\intp{}\frac{I(b\alu{1}{}-a\alp{0}{}|\vec{u}-\p{})}
{b\alu{1}{}-a\alp{0}{}}\,&\left[K_0^2\,\E^{0a}(\p{})-\frac{\ku{}{1b}}
{\ep_1}\,(\ku{}{1b}-\kp{}{0a})\cdot\E^{0a}(\p{})\right] \no \\
&=\frac{2\,b\,\alu{1}{}}{\,(\ep_1-\ep_0)}\,
\E^{1b}(\vec{u}) \, ,
\label{eqbP}
\end{align}
this expression represents in fact a set of two equations, depending on the
choice for $b = \pm$. 
The last step is to project \eref{eqbP} on the natural basis of
$\E^{1b}(\vec{u})$, namely $[\vec{u}]^{1b}\equiv (\evu{1b}{},\ehu{})$, 
which has 
the property to be orthogonal to $\ku{}{1b}$,  so it eliminates the
second term of \eref{eqbP} l.h.s. Let us notice that in order
to decompose $\E^{0a}(\p{})$ on $[\p{}]^{0a}$, one has to define a 
matrix $\op{M}^{1b,0a}(\vec{u}|\p{})$ transforming a vector 
expressed on the basis
$[\p{}]^{0a}$ into a vector on the basis $[\vec{u}]^{1b}$, multiplied by a
numerical factor $(\ep_0 \,\ep_1)^{\frac{1}{2}}\,K_0^2$ introduced for a
matter of convenience:
\begin{equation}
\op{M}^{1b,0a}(\vec{u}|\p{})\equiv (\ep_0\,\ep_1)^{\frac{1}{2}}\,K_0^2\,
\begin{pmatrix}
\evu{1b}{}\cdot\evp{0a}{} & \evu{1b}{}\cdot\ehp{} \\
\ehu{}\cdot\evp{0a}{} &\ehu{}\cdot\ehp{} \, .
\end{pmatrix}
\end{equation}

%%%%%%%%%%%%%%%%%%%%%%%%%%%%%%%%%%%%%%%%%%%%%%%%%%%%%%%%%%%%%%%%%%%%%%%%%%%%%%%
\subsection{The reduced Rayleigh equations}
%%%%%%%%%%%%%%%%%%%%%%%%%%%%%%%%%%%%%%%%%%%%%%%%%%%%%%%%%%%%%%%%%%%%%%%%%%%%%%%
With the definitions \sref{TE0}{TM1} the matrix $\op{M}$ takes the form:
\begin{equation}
\op{M}^{1b,0a}(\vec{u}|\p{})= \begin{pmatrix}
||\vec{u}||||\p{}||+ab\alu{1}{}\,\alp{0}{}\,\hvec{u}\cdot\hvec{p} & ~~~-b
\,\ep_0^{\frac{1}{2}}\,K_0\,\alu{1}{}\,(\hvec{u}\times\hvec{p})_z \\
a\,\ep_1^{\frac{1}{2}}\,K_0\,\alp{0}{}\,(\hvec{u}\times\hvec{p})_z &
(\ep_0\,\ep_1)^{\frac{1}{2}}\,K_0^2\,\hvec{u}\cdot\hvec{p}
\end{pmatrix}\, ,
\label{M10}
\end{equation}
and, the two reduced Raleigh equations resulting from \Eref{eqbP} read :
\begin{align}
\sum_{a=\pm}\intp{}\frac{I(b\alu{1}{}-a\alp{0}{}|\vec{u}-\p{})}{b\alu{1}{}-
a\alp{0}{}}\op{M}^{1b,0a}(\vec{u}|\p{})\,\E^{0a}(\p{})=\frac{2\,b\,
(\ep_0\,\ep_1)^{\frac{1}{2}}\,\alu{1}{}}{(\ep_1-\ep_0)}\,
\E^{1b}(\vec{u}) \, ,
\label{Reduced0}
\end{align}  
where we suppose that $\E^{0a}(\p{})$, $\E^{1b}(\vec{u})$
are respectively decomposed on the basis $[\p{}]^{0a}$ and $[\vec{u}]^{1b}$.
We can derive a similar equation where  $\E^{0b}$ is now eliminated, by simply
exchanging $\ep_0$ and $\ep_1$ in \eref{M10} and \eref{Reduced0},
due to the symetry of the equations \tref{DefE0}{DefE1},\sref{CCL1}{CCL3},
we get :
\begin{align}
\sum_{a=\pm}\intp{}\frac{I(b\alu{0}{}-a\alp{1}{}|\vec{u}-\p{})}{b\alu{0}{}-
a\alp{1}{}}\op{M}^{0b,1a}(\vec{u}|\p{})\,\E^{1a}(\p{})=
-\frac{2\,b\,(\ep_0\,\ep_1)^{\frac{1}{2}}\,\alu{0}{}}{
(\ep_1-\ep_0)}\, \E^{0b}(\vec{u})\, ,
\label{Reduced1}
\end{align}  
\begin{equation}
\op{M}^{0b,1a}(\vec{u}|\p{})=\begin{pmatrix}
||\vec{u}||||\p{}||+ab\alu{0}{}\,\alp{1}{}\,\hvec{u}\cdot\hvec{p} & ~~~-b
\,\ep_1^{\frac{1}{2}}\,K_0\,\alu{0}{}\,(\hvec{u}\times\hvec{p})_z \\
a\,\ep_0^{\frac{1}{2}}\,K_0\,\alp{1}{}\,(\hvec{u}\times\hvec{p})_z & 
(\ep_0\,\ep_1)^{\frac{1}{2}}\,K_0^2\,\hvec{u}\cdot\hvec{p}
\end{pmatrix}\, .
\label{M01}
\end{equation}
In the next sections, we will show how these equations greatly
simplify the perturbative calculation of plane waves scattering by
a rough surface.
In order to keep a compact notation, we introduce new matrices
$\op{M}_h$ given by :
\begin{align}
\op{M}_h^{1b,0a}(\vec{u}|\p{})&\equiv\frac{I(b\alu{1}{}-a\alp{0}{}|
\vec{u}-\p{})}{b\alu{1}{}-a\alp{0}{}}\op{M}^{1b,0a}(\vec{u}|\p{}) \, ,\\
\op{M}_h^{0b,1a}(\vec{u}|\p{})&\equiv\frac{I(b\alu{0}{}-a\alp{1}{}|
\vec{u}-\p{})}{b\alu{0}{}-a\alp{1}{}}\op{M}^{0b,1a}(\vec{u}|\p{}) \, .
\end{align}

%%%%%%%%%%%%%%%%%%%%%%%%%%%%%%%%%%%%%%%%%%%%%%%%%%%%%%%%%%%%%%%%%%%%%%%%%%%%%%%
\section{The diffusion matrix}
%%%%%%%%%%%%%%%%%%%%%%%%%%%%%%%%%%%%%%%%%%%%%%%%%%%%%%%%%%%%%%%%%%%%%%%%%%%%%%%
\label{Rmatrix}
We are interested by the diffusion of an incident plane wave by a
rough surface from the previous formalism. We define an
incident plane wave of wave vector $\kp{0}{0-}$ as:
\begin{equation}
\E^{0-}(\p{})=(2\pi)^2\,\delta(\p{}-\p{0})\E^{i}(\p{0})\, .
\end{equation}
We are naturally led to introduce the diffusion operator $\op R$:
\begin{eqnarray}
\vec{E}^{0+}(\p{} )\equiv \Rp{}{0}\cdot \vec{E}^{i}(\p 0) \, ,
\end{eqnarray}
which can be represented in a matrix form, using the vectorial basis
described above :
\begin{displaymath}
\Rp{}{0}=\begin{matrice} R_{VV}(\p{}|\p{0}) & R_{VH}(\p{}|\p{0}) \\
R_{HV}(\p{}|\p{0}) & R_{HH}(\p{}|\p{0})
\end{matrice}_{[p_0^{-}]\rightarrow [p^{+}]} \, .
\end{displaymath}
The field in medium 0 is now written (using the decomposition
\eref{DefE0}):
\begin{equation}
\vec{E}^{0}(\vec{r})=\vec{E}^{i}(\p{0})\e(\rmi\kp{0}{0-}\cdot\vec{r} ) 
+\intp{} \Rp{}{0}\cdot \vec{E}^{i}(\p 0)\e(\rmi\kp{}{0+}\cdot\vec{r} )
\label{DefEp}\, .
\end{equation}

%%%%%%%%%%%%%%%%%%%%%%%%%%%%%%%%%%%%%%%%%%%%%%%%%%%%%%%%%%%%%%%%%%%%%%%%%%%%%%%
\section{A perturbative development}
%%%%%%%%%%%%%%%%%%%%%%%%%%%%%%%%%%%%%%%%%%%%%%%%%%%%%%%%%%%%%%%%%%%%%%%%%%%%%%%
\label{Pertu}
In order to obtain a perturbative development, one has to make a
perturbative analysis of the given boundary-problem. A direct approach 
which uses an
exact integral equation named the extended boundary condition (EBC)
(see \citeo{Kong,Nieto}) requires tedious calculations. 
An other issue is to use the Rayleigh hypothesis in the boundary conditions. 
This is the method generally used  
to obtain (SPM)\cite{Rice,VoroSSA,Johnson}. But a great deal of
simplifications can be achieved if we are only interested by the
field outside the slab. It was discovered by Brown {\it et al.}\cite{Celli}, 
that an exact integral equation can obtained (excepted for the Rayleigh
hypothesis), which only involves the scattering matrix $\Rp{}{0}$.
The proof is based on the extinction theorem which decouples
the fields inside and outside the media. In this section, we will show how
to obtain this integral equation from the previous development, including
a generalization to the case of bounded random media.
%%%%%%%%%%%%%%%%%%%%%%%%%%%%%%%%%%%%%%
We seek for a perturbative development of $\op{R}$ in power of the height $h$:
\begin{equation}
\op{R}(\p{}|\p{0})=\op{R}^{(0)}(\p{}|\p{0})+\op{R}^{(1)}(\p{}|\p{0})+
\op{R}^{(2)}(\p{}|\p{0})+\op{R}^{(3)}(\p{}|\p{0})+\cdots \,.
\end{equation}
One can easily prove, that this development takes the following 
form (see \Aref{A2}):
\begin{eqnarray}
\op{R}(\p{}|\p{0})&=&(2\pi)^2\delta(\p{}-\p{0})\,\op{X}^{(0)}(\p{0})+
\alp{0}{0}\,
\op{X}^{(1)}(\p{}|\p{0})\, h(\p{}-\p{0}) \label{Dev1}\nonumber\\
& & +\alp{0}{0}\,
\intp{1}\,\op{X}^{(2)}(\p{}|\p{1}|\p{0})\, h(\p{}-\p{1})
h(\p{1}-\p{0}) \nonumber \\ 
& &+\alp{0}{0}\,
\iintp{1}{2} \,\op{X}^{(3)}(\p{}|\p{1}|\p{2}|\p{0})\, h(\p{}-\p{1})
h(\p{1}-\p{2})h(\p{2}-\p{0})  \, ,\label{Dev2}
\end{eqnarray}
where $h(\p{})$ is the Fourier transform\footnote{We use the same 
symbol for a function and its Fourier transform,
they are differentiated by their arguments} of $\hx$:
\begin{equation}
h(\p{})\equiv\int \rmd^2 \x\,\e(-i\p{}\cdot\x)\,\hx\, .
\end{equation}
We will now exemplify the power of the reduced Rayleigh equation
for the three configurations mentioned in the introduction.
%%%%%%%%%%%%%%%%%%%%%%%%%%%%%%%%%%%%%%%%%%%%%%%%%%%%%%%%%%%%%%%%%%%%%%%%%%%%%%%
\subsection{A rough surface separating two different media}
\begin{figure}[ht]
\input{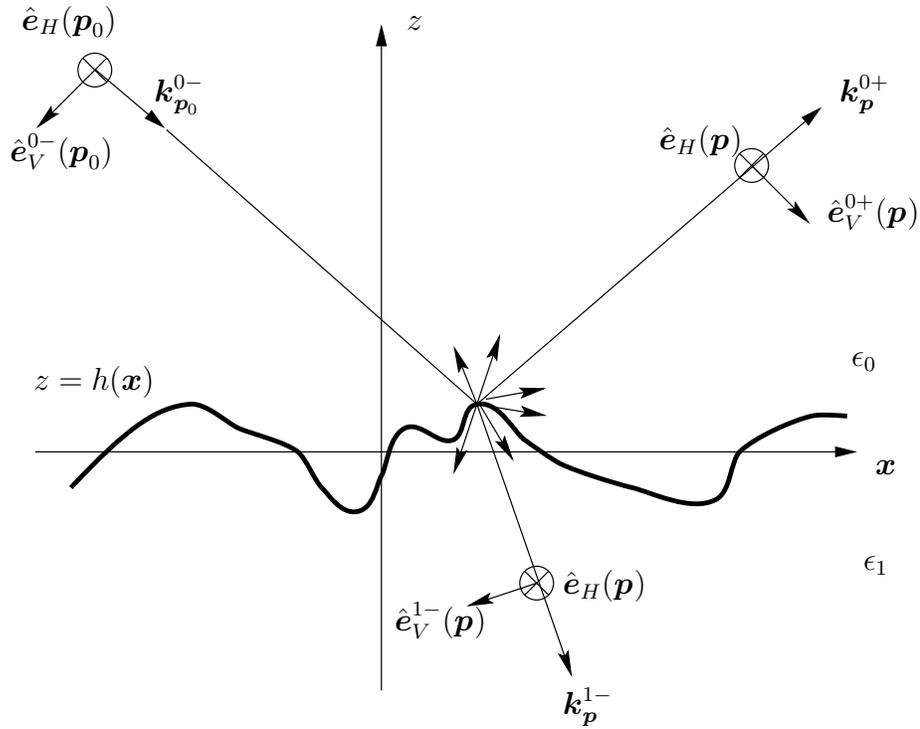}
\caption{A two-dimensional rough surface
separating two dielectric media 0 and 1.}
\label{surfaceSimple}
\end{figure}

%%%%%%%%%%%%%%%%%%%%%%%%%%%%%%%%%%%%%%%%%%%%%%%%%%%%%%%%%%%%%%%%%%%%%%%%%%%%%%%
We consider a rough surface delimiting two media which are semi infinite,
see \Fref{surfaceSimple}.
We suppose that there is no upward field propagating in the medium 1, 
so $\E^{1+}=0$.
With the choice, $b=+$, in equation \eref{Reduced0}, we obtain the
following integral equation for the scattering matrix
$\op{R}_{s\,\ep_0,\ep_1}(\p{}|\p{0})$ for a single surface\footnote{We 
show explicitely the permittivity dependance of $\op{R}_{s}$ with the 
subscript $\ep_0,\ep_1$ because in the following section we will use 
the $\op{R}_{s}$ matrix with different permittivity values.}
(the subscript $s$ means a single surface located at $z = 0$)
\begin{equation}
\intp{}\op{M}_h^{1+,0+}(\vec{u}|\p{})\cdot
\op{R}_{s\,\ep_0,\ep_1}(\p{}|\p{0})+\op{M}_h^{1+,0-}
(\vec{u}|\p{0})=0
\label{Reduced}
\end{equation}  
(This equation has been already obtained making use of the
extinction theorem.\cite{Celli} 
It has to be noticed that since the r.h.s of
\eref{Reduced} is null, one can simplify the second lign of the matrices,
$\op{M}^{1+,0+}$,$\op{M}^{1+,0-}$,  by a factor
$(\ep_1)^{\frac{1}{2}}$, then they 
coincide with the $\op{M}$,$\op{N}$ matrices derived by 
Celli {\it et al}~\cite{Celli}.)

In order to construct a perturbative development, the method is simply to expand 
in Taylor series the term $\e(i\alpha\,\hx)$ inside $I(\alpha|\p{})$, 
(\Eref{ialp}) :
\begin{align}
I(\alpha|\p{})&=(2\pi)^2\,\delta(\p{})-i\alpha\,h^{(1)}(\p{})-
\frac{\alpha^2}{2}\,h^{(2)}(\p{})-\frac{i\alpha^3}{3!}\,h^{(3)}(\p{})+\cdots 
\, ,\label{taylor} \\
h^{(n)}(\p{})&\equiv\int \rmd^2 \x\,\e(-i\p{}\cdot\x)\,h^n(\x) \, ,
\end{align}
and, to collect the terms of the same order in $\hx$.
Let us define the matrix
\begin{equation}
\op{D}_{10}^{\pm}(\p{0})\equiv\begin{pmatrix}
\ep_1\,\alp{0}{0}\pm\ep_0 \,\alp{1}{0} & 0\\
0 & \alp{0}{0}\pm\alp{1}{0} 
\end{pmatrix}\, , 
\label{D10}
\end{equation}
the classical specular reflection coefficients for $(TM)$ and $(TE)$
waves are given by the diagonal elements of the matrix
\begin{equation}
\op{V}^{10}(\p{0})\equiv\op{D}_{10}^{\,-}(\p{0})\cdot
\left[\op{D}_{10}^{+}(\p{0})
\right]^{-1} \, .\label{V10}
\end{equation} 
Introducing \eref{taylor} in \eref{Reduced}, we obtain for
$\op{R}_{s\,\ep_0,\ep_1}$ a perturbative
development of the form \eref{Dev1}, where the coefficients are given by:
\begin{align}
\op{X}_{s\,\ep_0,\ep_1}^{(0)}(\p{0})&=-\frac{\alp{1}{0}-\alp{0}{0}}
{\alp{1}{0}+\alp{0}{0}}\,[\op{M}^{1+,0+}(\p{0}|\p{0})]^{-1}\cdot\op{M}^{1+,0-}
(\p{0}|\p{0}) \no \\
&=\op{V}^{10}(\p{0}) \, ,
\end{align}   
and 
\begin{align}
\op{X}_{s\,\ep_0,\ep_1}^{(1)}(\vec{u}|\p{0})&=2i\,
\op{Q}^{+}(\vec{u}|\p{0}) \, ,\label{ordre1}\\
\op{X}_{s\,\ep_0,\ep_1}^{(2)}(\vec{u}|\p{1}|\p{0})&=\alu{1}{}\,
\op{Q}^{+}(\vec{u}|\p{0})+
\alp{0}{0}\,\op{Q}^{\,-}(\vec{u}|\p{0})-2\,\op{P}(\vec{u}|\p{1})
\cdot\op{Q}^{+}(\p{1}|\p{0})
\label{ordre2} \, ,\\
\op{X}_{s\,\ep_0,\ep_1}^{(3)}(\vec{u}|\p{1}|\p{2}|\p{0})&=-\frac{i}{3}
\left[(\alpha_1^{2}(\vec{u})+\alpha_0^2(\p{0}))\,\op{Q}^+(\vec{u}|\p{0})+2\,
\alu{1}{}\,\alp{0}{0}\,\op{Q}^{\,-}(\vec{u}|\p{0})\right] \no \\
&+i\,\op{P}(\vec{u}|\p{1})\,\op{X}^{(2)}_{s\,\ep_0,\ep_1}(\p{1}|\p{2}|\p{0})+i\,
(\alu{1}{}-\alp{0}{2})\cdot\op{P}(\vec{u}|\p{2})\cdot\op{Q}^{+}(\p{2}|\p{0}) 
\, ,\label{ordre3}
\end{align}
 with
\begin{align}
\op{Q}^{\pm}(\vec{u}|\p{0})&\equiv\frac{\alu{1}{}-\alu{0}{}}{2\,\alp{0}{0}}\,
[\op{M}^{1+,0+}(\vec{u}|\vec{u})]^{-1}\cdot[\op{M}^{1+,0-}(\vec{u}|\p{0})\pm
\op{M}^{1+,0+}(\vec{u}|\p{0})\cdot 
\op{X}^{(0)}(\p{0})]\, ,\\
\op{P}(\vec{u}|\p{1})&\equiv
(\alu{1}{}-\alu{0}{})\,[\op{M}^{1+,0+}(\vec{u}|\vec{u})]^{-1}\cdot\op{M}^{1+,0+}
(\vec{u}|\p{1})]
\end{align}
and, after some simple algebra we obtain:
\begin{align}
\op{Q}^{+}(\vec{u}|\p{0})&=(\ep_1-\ep_0)\,[\op{D}_{10}^{+}(\vec{u})]^{-1}\cdot
\begin{pmatrix}
\ep_1\,||\vec{u}||||\p{0}||-\ep_0\,\alu{1}{}\,\alp{1}{0}\,\hvec{u}
\cdot\hvec{p}_0 &
-\ep_0^{\frac{1}{2}}\,K_0\,\alu{1}{}\,(\hvec{u}\times\hvec{p}_0)_z\\
-\ep_0^{\frac{1}{2}}\,K_0\,\alp{1}{0}\,(\hvec{u}\times\hvec{p}_0)_z
&K_0^2\,\hvec{u}\cdot\hvec{p}_0
\end{pmatrix} \no \\
&\cdot[\op{D}_{10}^{+}(\p{0})]^{-1} \, ,\label{Q+}\\
\op{Q}^{~-}(\vec{u}|\p{0})
&=\frac{(\ep_1-\ep_0)}{\alp{0}{0}}\,[\op{D}_{10}^{+}(\vec{u})]^{-1}\,\cdot 
\no \\ &\begin{pmatrix}
\ep_0\,\alp{1}{0}\,||\vec{u}||||\p{0}||-\ep_1\,\alu{1}{}\,\alpha_0^2(\p{0})\,
\hvec{u}\cdot\hvec{p}_0 &
-\ep_0^{\frac{1}{2}}\,K_0\,\alu{1}{}\,\alp{1}{0}\,(\hvec{u}\times\hvec{p}_0)_z\\
-\ep_0^{-\frac{1}{2}}\,\ep_1\,K_0\,\alpha_0^2(\p{0})\,
(\hvec{u}\times\hvec{p}_0)_z
&K_0^2\,\alp{1}{0}\,\hvec{u}\cdot\hvec{p}_0
\end{pmatrix}
\cdot[\op{D}_{10}^{+}(\p{0})]^{-1} \, ,\label{Q-}
\end{align}   
\begin{align}
\op{P}(\vec{u}|\p{1})&=(\ep_1-\ep_0)\,[\op{D}_{10}^{+}(\vec{u})]^{-1}\cdot
\begin{pmatrix}
||\vec{u}||||\p{1}||+\alu{1}{}\,\alp{0}{1}\,\hvec{u}\cdot\hvec{p}_1 &
-\ep_0^{\frac{1}{2}}\,K_0\,\alu{1}{}\,(\hvec{u}\times\hvec{p}_1)_z \\
\ep_0^{-\frac{1}{2}}\,K_0\,\alp{0}{1}\,(\hvec{u}\times\hvec{p}_1)_z & K_0^2\,
\hvec{u}\cdot\hvec{p}_1
\end{pmatrix}\, .\label{P}
\end{align}  
It can be easily checked that $\op{X}^{(1)}$ is the well known first order
term in perturbation theory which was obtained by Rice.~\cite{Rice}
After some lenghty calculations, we have proven that
\Tref{ordre2}{ordre3}, are identical to those found by Johnson.~\cite{Johnson} 
Thus our expressions \eref{ordre2}, \eref{ordre3}, are a compact manner to 
write the second and third order terms of the perturbative expansion,
moreover, they are well adapted for numerical computations.
However, it has to be noticed that only the first term $\op{X}^{(1)}$
is reciprocal. Since the second and third-order perturbative terms are
included in an integral, the coefficient $\op{X}^{(2)}$, $\op{X}^{(3)}$,
are not unique, however they can be put into a reciprocal form (see \Aref{A2}).

It is worth to notice that we can follow an analogous procedure to calculate 
the transmitted field. By taking $b=-$ in  \Eref{Reduced1}, we get:
\begin{equation}
\intp{}\op{M}_h^{0-,1-}(\vec{u}|\p{})\cdot\E^{1-}(\p{})=\frac{2\,
(\ep_0\,\ep_1)^{\frac{1}{2}}\,\alu{0}{}}{(\ep_1-\ep_0)}\,
\E^{0-}(\vec{u}) \, .
\end{equation}
This equation was already obtained  with the extinction theorem.~\cite{Greffet}

%%%%%%%%%%%%%%%%%%%%%%%%%%%%%%%%%%%%%%%%%%%%%%%%%%%%%%%%%%%%%%%%%%%%%%%%%%%%%%%
\subsection{A slab with a rough surface on the bottom side}
%%%%%%%%%%%%%%%%%%%%%%%%%%%%%%%%%%%%%%%%%%%%%%%%%%%%%%%%%%%%%%%%%%%%%%%%%%%%%%%

\begin{figure}[ht]
\input{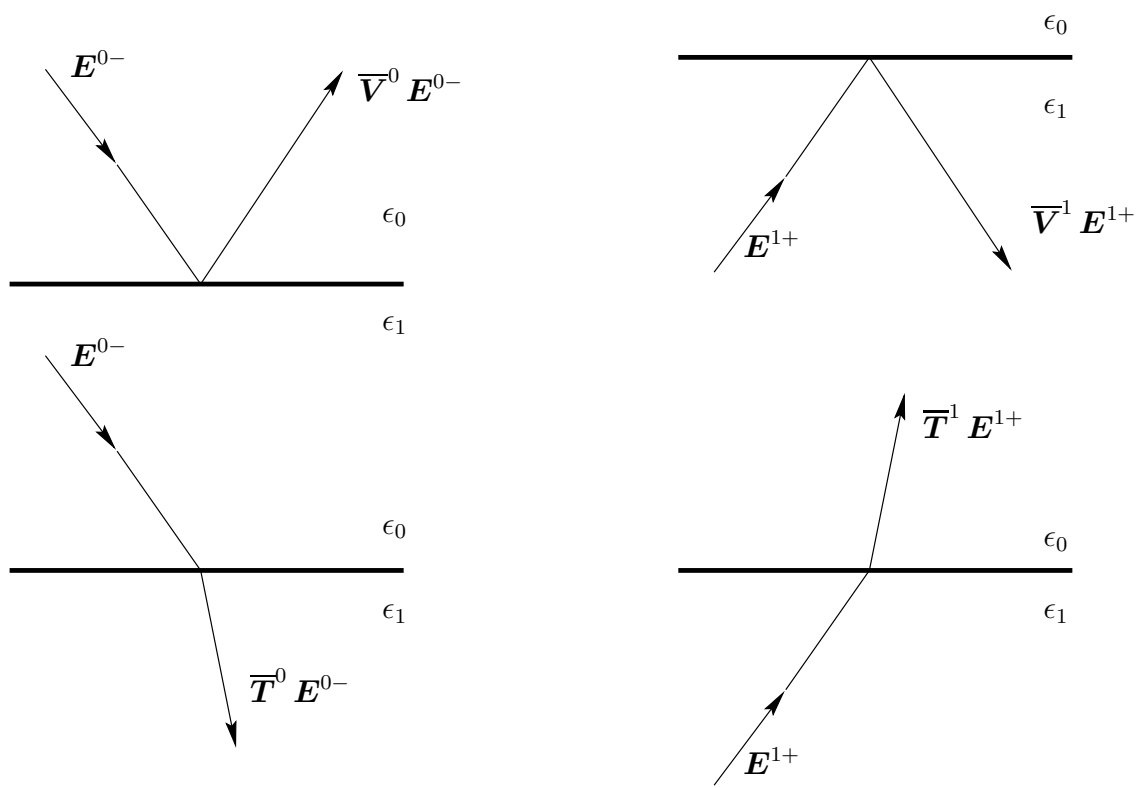}
\caption{Definitions of scattering matrices for a planar surface.}
\label{Defplane}
\end{figure}

\begin{figure}[ht]
% figure 6
\input{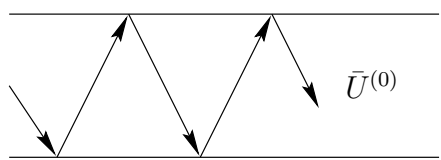}
\caption{Diagrammatic representation of the operator $\op{U}^{(0)}$.}
\label{MatU0}
\end{figure}

We consider a slab delimited on the upper side by a planar surface and on the
bottom side by a rough surface, see \Fref{slabB}. Since there is
no incident upward field in medium 2, the scattering matrix
obtained in the preceeding section is sufficient to determine the
scattering matrix of the present configuration.
In order to get a proof, let us introduce some definitions
as explained in \Fref{Defplane}.
The scattering matrix for an incident plane wave coming from the medium 0, 
and scattered in the medium 1 is given by:
\begin{equation}
\op{V}^{0}(\p{}|\p{0})=(2\pi)^2\,\delta(\p{}-\p{0})\,\op{V}^{10}(\p{0})\, 
\label{V0} ,
\end{equation}
where $\op{V}^{10}$ is defined by \eref{V10}.
The transmitted wave in the medium 1 is given by:
\begin{align}
\op{T}^{0}(\p{}|\p{0})&=(2\pi)^2\,\delta(\p{}-\p{0})\,
\frac{\alp{0}{0}}{\alp{1}{0}}\,\op{T}^{10}(\p{0})\, , \label{T0}\\
\op{T}^{10}(\p{0})&\equiv2\alp{1}{0}\begin{pmatrix}
(\ep_0\,\ep_1)^{\frac{1}{2}}
& 0 \\
0 & 1
\end{pmatrix}\, .[\op{D}_{10}^{+}(\p{0})]^{-1} \, .
\end{align}
Now, when the incident wave is coming from the medium 1, we have similarly :
\begin{equation}
\op{V}^{1}(\p{}|\p{0})=-(2\pi)^2\,\delta(\p{}-\p{0})\,\op{V}^{10}(\p{0})\, , 
\label{V1}
\end{equation}
\begin{equation}
\op{T}^{1}(\p{}|\p{0})=(2\pi)^2\,\delta(\p{}-\p{0})\,\op{T}^{10}(\p{0})\, .
\label{T1}
\end{equation}
The scattering matrix $\op{R}^{H}_{s\,\ep_1,\ep_2}$ for the rough surface 
$h$ which is located at $z=-H$~\footnote{Throughout the paper, the
symbol $H$ in upper index is related to the height of a surface, 
in lower index to the polarization.}, and separating two media of
permittivity $\ep_1$ and $\ep_2$ is given by:
\begin{equation}
\op{R}^{H}_{s\,\ep_1,\ep_2}(\p{}|\p{0})=\e(\rmi(\alp{1}{}+\alp{1}{0})\,H)\,
\op{R}_{s\,\ep_1,\ep_2}(\p{}|\p{0})\, ,
\end{equation}
where the phase term comes from the translation $z=-H$, 
(see \eref{translation}),
and $\op{R}_{s\,\ep_1,\ep_2}$ denotes the scattering matrix
$\op{R}_{s\,\ep_0,\ep_1}$ of the previous section, where we have replaced 
$\ep_0$
by $\ep_1$, and $\ep_1$ by $\ep_2$.
Furthermore, if we define the product of two operator $\op{A}$ and
$\op{B}$ by:
\begin{equation}
(\op{A}\cdot\op{B})(\p{}|\p{0})\equiv\intp{1}\,
\op{A}(\p{}|\p{1})\cdot\op{B}(\p{1}|\p{0})\, ,
\end{equation}
one can easily prove for the configuration shown in \Fref{slabB} that
(we use for the fields the notations of \Fref{surface}), 
\begin{align}
\E^{1+}&=\op{R}^{H}_{s\,\ep_1,\ep_2}\cdot\op{T}^{0}\cdot\E^{0-}+
\op{R}^{H}_{s\,\ep_1,\ep_2}\cdot\op{V}^1\cdot\E^{1+} \, , 
\label{lip1}\\
\E^{0+}&=\op{V}^0\cdot\E^{0-}+\op{T}^1\cdot\E^{1+} \, ,\label{lip2}
\end{align}
where $\E^{0-}(\p{})=(2\pi)^2\,\delta(\p{}-\p{0})\E^{i}(\p{0})$.
These equations have been recently used to calculate in 
the first order the field scattered by a layered medium.~\cite{Fuks} 
In fact, as we shall see below, these equations allow us to
obtain all orders of the field perturbation. 
The expression \eref{lip1} is analogous to the Dyson equation usually used
in random media.~\cite{Kong} So, we are naturally led to introduce a
scattering operator $\op{U}$ :
\begin{equation}
\E^{1+}=\op{R}^{H}_{s\,\ep_1,\ep_2}\cdot\op{U}\cdot\op{T}^0\cdot\E^{0-} \, , 
\label{Dyson}
\end{equation}
which satisfies the equation
\begin{equation}
\op{U}=\op{1}+\op{V}^1\cdot\op{R}^{H}_{s\,\ep_1,\ep_2}\cdot\op{U} \,. \label{U}
\end{equation} 
If we define by $\op{R}_d(\p{}|\p{0})$ the global scattering matrix for 
the upper planar surface and the bottom rough surface by, 
\begin{equation}
\E^{0+}(\p{})=\op{R}_{d}(\p{}|\p{0})\cdot\E^{i}(\p{0})\, ,
\end{equation}
with \Tref{lip2}{Dyson}, the scattering matrix becomes:
\begin{equation}
\op{R}_{d}=\op{V}^0+\op{T}^1\cdot\op{R}^{H}_{s\,\ep_1,\ep_2}\cdot\op{U}
\cdot\op{T}^0 \, .
\label{Ru}
\end{equation}
We can improve the development \eref{U}, by summing all the specular
reflexions inside the slab, this can be done by introducing the operator
$\op{U}^{(0)}$ which satisfies the equation
\begin{equation}
\op{U}^{(0)}=\op{1}+\op{V}^1\cdot\op{R}^{H\,(0)}_{s\,\ep_1,\ep_2}
\cdot\op{U}^{(0)} \,, 
\label{U0}
\end{equation} 
where $\op{R}^{H\,(0)}_{s\,\ep_1,\ep_2}$ is the zeroth order term of the 
perturbative development which is given by:
\begin{align}
\op{R}^{H\,(0)}_{s\,\ep_1,\ep_2}(\p{}|\p{0})&=(2\pi)^2\,\delta(\p{}-\p{0})\,
\op{V}^{H\,21}(\p{0}) \, ,
\end{align}
and, $\op{V}^{H\, 21}$ is the scattering matrix for a planar surface located at
the height $z=-H$
\begin{align}
\op{V}^{H\,21}(\vec{u})&\equiv \exp(2\,i\,\alu{1}{}\,H)\,
\op{D}_{21}^{\,-}(\p{0})\cdot\left[\op{D}_{21}^{+}(\p{0})\right]^{-1} \, , 
\label{V21a}\\
\op{D}_{21}^{\pm}(\p{0})&\equiv\begin{pmatrix}
\ep_2\,\alp{1}{0}\pm\ep_1 \,\alp{2}{0} & 0\\
0 & \alp{1}{0}\pm\alp{2}{0} 
\end{pmatrix}\, .\label{V21}
\end{align}
The term $\e(2\,i\,\alu{1}{}\,H)$ comes from the phase shift induced by the
translation of the planar surface from the height $z=0$ to $z=-H$
(see \Aref{translation}).
The diagrammatic representation of \Eref{U0} is shown in \Fref{MatU0},
it is in fact a geometric series which can be summed
\begin{align}
\op{U}^{(0)}(\p{}|\p{0})&=(2\pi)^2\,\delta(\p{}-\p{0})\,\op{U}^{(0)}(\p{0})
\, ,\\
\op{U}^{(0)}(\p{0})&\equiv
\left[ \op{1}+\op{V}^{10}(\p{0})\cdot\op{V}^{H\,21}(\p{0})\right]^{-1}\, .
\end{align}
From the previous results, \Eref{U} can be written in the following
form :
\begin{align}
\op{U}&=\op{U}^{(0)}+\op{U}^{(0)}\cdot\op{V}^1\cdot\Delta
\op{R}^{H}_{s\,\ep_1,\ep_2}\cdot\op{U}\, , 
\label{Ur}
\end{align} 
where
\begin{equation}
\Delta\op{R}^{H}_{s\,\ep_1,\ep_2}\equiv\op{R}^{H}_{s\,\ep_1,\ep_2}-
\op{R}^{H\,(0)}_{s\,\ep_1,\ep_2}\, .
\end{equation}
In order to obtain the perturbative development of $\op{R}_d$, we
introduce the expansion
\begin{equation}
\op{R}^{H}_{s\,\ep_1,\ep_2}=\op{R}^{H\,(0)}_{s\,\ep_1,\ep_2}+
\op{R}^{H\,(1)}_{s\,\ep_1,\ep_2}+
\op{R}^{H\,(2)}_{s\,\ep_1,\ep_2}+\op{R}^{H\,(3)}_{s\,\ep_1,\ep_2}\, ,
\end{equation}
in \eref{Ru} and \eref{Ur}, which gives the following terms 
\begin{align}
\op{R}^{(0)}_d &=\op{V}^{0}+\op{T}^1\cdot\op{R}^{H\,(0)}_{s\,\ep_1,\ep_2}
\cdot\op{U}^{(0)}\cdot
\op{T}^0 \, ,\\
\op{R}^{(1)}_d &=\op{T}^{1}\cdot\op{U}^{(0)}
\cdot\op{R}^{H\,(1)}_{s\,\ep_1,\ep_2}\cdot
\op{U}^{(0)}\cdot\op{T}^{0}\, ,\\
\op{R}^{(2)}_d &=\op{T}^{1}\cdot\op{U}^{(0)}\cdot
\left[\op{R}^{H\,(2)}_{s\,\ep_1,\ep_2}
+\op{R}^{H\,(1)}_{s\,\ep_1,\ep_2}\cdot\op{U}^{(0)}\cdot\op{V}^{1}
\cdot \op{R}^{H\,(1)}_{s\,\ep_1,\ep_2}\right]\cdot\op{U}^{(0)}
\cdot\op{T}^{0}\,,\\
\op{R}^{(3)}_d &=\op{T}^{1}\cdot\op{U}^{(0)}\cdot
\left[\op{R}^{H\,(3)}_{s\,\ep_1,\ep_2}
+\op{R}^{H\,(2)}_{s\,\ep_1,\ep_2}\cdot\op{U}^{(0)}\cdot\op{V}^{1}
\cdot \op{R}^{H\,(1)}_{s\,\ep_1,\ep_2}+
\op{R}^{H\,(1)}_{s\,\ep_1,\ep_2}\cdot\op{U}^{(0)}\cdot\op{V}^{1}
\cdot \op{R}^{H\,(2)}_{s\,\ep_1,\ep_2} \right.\no \\ 
&\left.+\op{R}^{H\,(1)}_{s\,\ep_1,\ep_2}\cdot
\op{U}^{(0)}\cdot\op{V}^{1}
\cdot \op{R}^{H\,(1)}_{s\,\ep_1,\ep_2}\cdot\op{U}^{(0)}\cdot\op{V}^{1}
\cdot \op{R}^{H\,(1)}_{s\,\ep_1,\ep_2}\right]
\cdot\op{U}^{(0)}\cdot\op{T}^{0}\,.
\end{align}
Using the development \eref{Dev1} for $\op{R}^{H}_{s\,\ep_1,\ep_2}$, 
and the definitions \sref{V0}{T0}, \sref{V1}{T1},
we obtain after some calculations  a development of the form
\eref{Dev1} for $\op{R}_d$ with the following coefficients:
\begin{equation}
\op{X}^{(0)}_d(\p{0})=\left(\op{V}^{10}(\p{0})+\op{V}^{H\,21}(\p{0})\right)
\cdot\left[\op{1}+\op{V}^{10}(\p{0})\cdot\op{V}^{H\,21}(\p{0})\right]^{-1} \, ,
\end{equation}
this matrix is naturally diagonal, and its coefficients are identical to those
of the reflection coefficients for a planar slab~\cite{Kong}.
The other coefficients are
\begin{align}
\op{X}^{(1)}_d(\p{}|\p{0})&=\op{T}^{10}(\p{})\cdot\op{U}^{(0)}(\p{})\cdot
\op{X}_{s\,\ep_1,\ep_2}^{H\,(1)}(\p{}|\p{0})\cdot\op{U}^{(0)}(\p{0})\cdot
\op{T}^{10}(\p{0})\, ,\label{Xb1}\\
\op{X}^{(2)}_d(\p{}|\p{1}|\p{0})&=\op{T}^{10}(\p{})\cdot\op{U}^{(0)}(\p{})\cdot
\left[\op{X}_{s\, 
\ep_1,\ep_2}^{H\,(2)}(\p{}|\p{1}|\p{0})\right.\no \\
&\left.-\alp{1}{1}\,\op{X}_{s\, \ep_1,\ep_2}^{H\,(1)}
(\p{}|\p{1})\cdot\op{U}^{(0)}(\p{1})\cdot\op{V}^{10}(\p{1})
\cdot \op{X}_{s\,\ep_1,\ep_2}^{H\,(1)}(\p{1}|\p{0})\right]
\cdot\op{U}^{(0)}(\p{0})\cdot\op{T}^{10}(\p{0})\,,\label{X2d}\\
\op{X}^{(3)}_d(\p{}|\p{1}|\p{2}|\p{0})&=\op{T}^{10}(\p{})\cdot
\op{U}^{(0)}(\p{})\cdot\left[\op{X}_{s \, \ep_1,\ep_2}^{H\,(3)}
(\p{}|\p{1}|\p{2}|\p{0})\right.\no\\
&-\alp{1}{2}\,\op{X}_{s\,\ep_1,\ep_2}^{H\,(2)}(\p{}|\p{1}|\p{2})
\cdot\op{U}^{(0)}(\p{2})\cdot\op{V}^{10}(\p{2})
\cdot \op{X}_{s\,\ep_1,\ep_2}^{H\,(1)}(\p{2}|\p{0})\no\\
& -\alp{1}{1}\,\op{X}_{s\,\ep_1,\ep_2}^{H\,(1)}(\p{}|\p{1})
\cdot\op{U}^{(0)}(\p{1})\cdot\op{V}^{10}(\p{1})
\cdot \op{X}_{s\,\ep_1,\ep_2}^{H\,(2)}(\p{1}|\p{2}|\p{0})\no\\
& +\alp{1}{1}\,\alp{1}{2}\,\op{X}_{s\,\ep_1,\ep_2}^{H\,(1)}(\p{}|\p{1})
\cdot\op{U}^{(0)}(\p{1})\cdot\op{V}^{10}(\p{1})
\cdot \op{X}_{s\,\ep_1,\ep_2}^{H\,(1)}(\p{1}|\p{2})\no \\
&\left.\cdot\op{U}^{(0)}(\p{2})\cdot\op{V}^{10}(\p{2})\cdot 
\op{X}_{s\,\ep_1,\ep_2}^{H\,(1)}(\p{2}|\p{0})\right]
\cdot\op{U}^{(0)}(\p{0})\cdot\op{T}^{10}(\p{0})\,.\label{Xb3}
\end{align}
In these expressions,
$\op{X}_{s\,\ep_1,\ep_2}^{H\,(n)}(\p{}|\p{0})\equiv\e(\rmi(\alp{1}{}+
\alp{1}{0})\,H)\,\op{X}_{s\,\ep_1,\ep_2}^{(n)}(\p{}|\p{0})$, 
and the subscripts $\ep_1,\,\ep_2$ in
$\op{X}_{s\,\ep_1,\ep_2}$ means that we replace $\ep_0$ by $\ep_1$,
and  $\ep_1$ by $\ep_2$ in  \sref{ordre1}{ordre3}.

%%%%%%%%%%%%%%%%%%%%%%%%%%%%%%%%%%%%%%%%%%%%%%%%%%%%%%%%%%%%%%%%%%%%%%%%%%%%%%%
\subsection{A slab with a rough surface on the upper side}
%%%%%%%%%%%%%%%%%%%%%%%%%%%%%%%%%%%%%%%%%%%%%%%%%%%%%%%%%%%%%%%%%%%%%%%%%%%%%%%
\label{slabproug}
We consider a slab delimited on the upper side by a two-dimensional rough
surface, and on the bottom side by a planar surface, see \Fref{slabU}.
To derive the reduced Rayleigh equation for this configuration,
we have to combine the two following equations :
\begin{align}
\intp{} \op{M}_h^{1+,0+}(\vec{u}|\p{})\cdot\op{R}_u(\p{}|\p{0})\cdot
\E^{i}(\p{0})+
\op{M}_h^{1+,0-}(\vec{u}|\p{0})\cdot\E^{i}(\p{0})=\frac{2\,
(\ep_0\,\ep_1)^{\frac{1}{2}}\,\alu{1}{}}{(\ep_1-\ep_0)}\,
\E^{1+}(\vec{u}) \, ,\\
\intp{}\op{M}_h^{1-,0+}(\vec{u}|\p{})\cdot\op{R}_u(\p{}|\p{0})\cdot
\E^{i}(\p{0})+
\op{M}_h^{1-,0-}(\vec{u}|\p{0})\cdot\E^{i}(\p{0})=-\frac{2\,
(\ep_0\,\ep_1)^{\frac{1}{2}}\,\alu{1}{}}{(\ep_1-\ep_0)}\,
\E^{1-}(\vec{u}) \, ,
\end{align}  
with 
\begin{equation}
\E^{1+}(\vec{u})=\op{V}^{H\, 21}(\vec{u})\cdot\E^{1-}(\vec{u}) \, ,
\end{equation}
where $\op{V}^{H\, 21}$ is given by \eref{V21}, and $\op{R}_u$ is the
global scattering matrix for the upper rough surface and the bottom
planar surface.

The reduced Rayleigh equation for the scattering matrix $\op{R}_{u}$ is then 
\begin{align}
\intp{}\,&
\left[\op{M}_h^{1+,0+}(\vec{u}|\p{})+
\op{V}^{H\,21}(\vec{u})\cdot\op{M}_h^{1-,0+}(\vec{u}|\p{})\right]\cdot
\op{R}_u(\p{}|\p{0})=\no \\
&-\left[\op{M}_h^{1+,0-}(\vec{u}|\p{0})+\op{V}^{H\,21}(\vec{u})\cdot
\op{M}_h^{1-,0-}(\vec{u}|\p{0})\right]\, .
\end{align}
With the expansion of $I(\alpha|\p{})$ in power series, we obtain the
perturbative development:
\begin{align}
\op{X}_{u}^{(0)}(\p{0})&=-\left[\frac{\op{M}^{1+,0+}
(\p{0}|\p{0})}{\alp{1}{0}-
\alp{0}{0}}-\op{V}^{H\,21}(\p{0})\cdot\frac{\op{M}^{1-,0+}(\p{0}|\p{0})}
{\alp{1}{0}+\alp{0}{0}}\right]^{-1} \no \\
&~~~~~~\cdot\left[\frac{\op{M}^{1+,0-}(\p{0}|\p{0})}{\alp{1}{0}+\alp{0}{0}}+
\op{V}^{H\,21}(\p{0})\cdot\frac{\op{M}^{1-,0-}(\p{0}|\p{0})}
{-\alp{1}{0}+\alp{0}{0}}\right] 
\no\\
&=\left(\op{V}^{10}(\p{0})+\op{V}^{H\,21}(\p{0})\right)
\cdot\left[\op{1}+\op{V}^{10}(\p{0})\cdot\op{V}^{H\,21}(\p{0})\right]^{-1} \, ,
\end{align}

\begin{align}
\op{X}_{u}^{(1)}(\vec{u}|\p{0})&\equiv2\,i\,
\op{Q}^{++}(\vec{u}|\p{0})
\label{ordre1slab}\\
\op{X}_{u}^{(2)}(\vec{u}|\p{1}|\p{0})&=\alu{1}{}\,
\op{Q}^{\,-+}(\vec{u}|\p{0})+
\alp{0}{0}\,\op{Q}^{+-}(\vec{u}|\p{0})-2\,\op{P}^+(\vec{u}|\p{1})\cdot
\op{Q}^{++}(\p{1}|\p{0})
\label{ordre2slab}\\
\op{X}_{u}^{(3)}(\vec{u}|\p{1}|\p{2}|\p{0})
&=-\frac{i}{3}\left[(\alpha_1^{2}
(\vec{u})+\alpha_0^2(\p{0}))\,\op{Q}^{++}(\vec{u}|\p{0})+2\,\alu{1}{}\,
\alp{0}{0}\,\op{Q}^{\,--}(\vec{u}|\p{0})\right] \no \\
&+i\,\op{P}^{+}(\vec{u}|\p{1})\cdot\op{X}^{(2)}(\p{1}|\p{2}|\p{0})+i\,
\left[\alu{1}{}\,\op{P}^{-}(\vec{u}|\p{2})-\alp{0}{2}\,\op{P}^{+}
(\vec{u}|\p{2})\right]\cdot\op{Q}^{++}(\p{2}|\p{0})
\label{ordre3slab}
\end{align}
with 
\begin{align}
\op{Q}^{ba}(\vec{u}|\p{0})\equiv&\frac{1}{2\,\alp{0}{0}}\,
\left[\frac{\op{M}^{1+,0+}(\vec{u}|\vec{u})}{\alu{1}{}-\alu{0}{}}-
\op{V}^{H\,21}(\vec{u})\cdot\frac{\op{M}^{1-,0+}(\vec{u}|\vec{u})}
{\alu{1}{}+\alu{0}{}}\right]^{-1}\no \\
&\cdot\left[a\,\op{M}^{1+,0+}(\vec{u}|\p{0})\cdot
\op{X}_{s\,\ep_0,\ep_1}^{(0)}(\p{0})+
\op{M}^{1+,0-}(\vec{u}|\p{0})\right. \no \\
& \left.+b\,\op{V}^{H\,21}(\vec{u})\cdot\left(a\,\op{M}^{1-,0+}(\vec{u}|\p{0})\cdot
\op{X}_{s\,\ep_0,\ep_1}^{(0)}(\p{0})+\op{M}^{1-,0-}(\vec{u}|\p{0})\right)
\right]\, ,\\
\op{P}^{\pm}(\vec{u}|\p{1})\equiv& \left[\frac{\op{M}^{1+,0+}(\vec{u}|\vec{u})}
{\alu{1}{}-\alu{0}{}}-
\op{V}^{H\,21}(\vec{u})\cdot\frac{\op{M}^{1-,0+}(\vec{u}|\vec{u})}
{\alu{1}{}+\alu{0}{}}\right]^{-1} \no \\
& \cdot
\left[\op{M}^{1+,0+}(\vec{u}|\p{1})\pm\op{V}^{H\,21}(\vec{u})\cdot\op{M}^{1-,0+}
(\vec{u}|\p{1})\right]\, ,
\end{align}
where, $a=\pm$, $b=\pm$ are the sign indices.
After some computations we obtain:
\begin{align}
&\op{Q}^{++}(\vec{u}|\p{0})=(\ep_1-\ep_0)\,[\op{D}_{10}^{+}(\vec{u})]^{-1}\cdot 
\no \\
&\quad \begin{pmatrix}
\ep_1\,||\vec{u}||||\p{0}||\,F_V^+(\vec{u})F_V^+(\p{0})&
-\ep_0^{\frac{1}{2}}\,K_0\,\alu{1}{}\,F_V^-(\vec{u})\,F_H^+(\p{0})
(\hvec{u}\times\hvec{p}_0)_z\\
-\ep_0\,\alu{1}{}\,
\alp{1}{0}\,F_V^-(\vec{u})\,F_V^-(\p{0})\,\hvec{u}\cdot\hvec{p}_0
 & \\ & \\
-\ep_0^{\frac{1}{2}}\,K_0\,\alp{1}{0}\,F_H^+(\vec{u})\,F_V^-(\p{0})\,
(\hvec{u}\times\hvec{p}_0)_z
&K_0^2\,F_H^+(\vec{u})\,F_H^+(\p{0})\,\hvec{u}\cdot\hvec{p}_0
\end{pmatrix}\cdot [\op{D}_{10}^{+}(\p{0})]^{-1}\, ,
\label{Q++}
\end{align}
\begin{align}
&\op{Q}^{-+}(\vec{u}|\p{0})=(\ep_1-\ep_0)\,[\op{D}_{10}^{+}(\vec{u})]^{-1}\cdot 
\no \\
&\quad \begin{pmatrix}
\ep_1\,||\vec{u}||||\p{0}||\,F_V^-(\vec{u})F_V^+(\p{0}) &
-\ep_0^{\frac{1}{2}}\,K_0\,\alu{1}{}\,F_V^+(\vec{u})\,F_H^+(\p{0})
(\hvec{u}\times\hvec{p}_0)_z\\
-\ep_0\,\alu{1}{}\,
\alp{1}{0}\,F_V^+(\vec{u})\,F_V^-(\p{0})\,\hvec{u}\cdot\hvec{p}_0 & \\
& \\
-\ep_0^{\frac{1}{2}}\,K_0\,\alp{1}{0}\,F_H^-(\vec{u})\,F_V^-(\p{0})\,
(\hvec{u}\times\hvec{p}_0)_z
&K_0^2\,F_H^-(\vec{u})\,F_H^+(\p{0})\,\hvec{u}\cdot\hvec{p}_0
\end{pmatrix} \cdot [\op{D}_{10}^{+}(\p{0})]^{-1} \, ,
\label{Q-+}
\end{align}
\begin{align}
&\op{Q}^{+-}(\vec{u}|\p{0})
=\frac{(\ep_1-\ep_0)}{\alp{0}{0}}\,[\op{D}_{10}^{+}(\vec{u})]^{-1}\, \cdot \no
\\ &\begin{pmatrix}
\ep_0\,\alp{1}{0}\,||\vec{u}||||\p{0}||F_V^+(\vec{u})F_V^-(\p{0}) &
-\ep_0^{\frac{1}{2}}\,K_0\,\alu{1}{}\,\alp{1}{0}\,F_V^-(\vec{u})F_H^-(\p{0})\,
(\hvec{u}\times\hvec{p}_0)_z\\
-\ep_1\,
\alu{1}{}\,\alpha_0^2(\p{0})\,F_V^-(\vec{u})F_V^+(\p{0})\,\hvec{u}\cdot
\hvec{p}_0 & \\
& \\
-\ep_0^{-\frac{1}{2}}\,\ep_1\,K_0\,\alpha_0^2(\p{0})\,F_H^+(\vec{u})F_V^+(\p{0})
\, (\hvec{u}\times\hvec{p}_0)_z
&K_0^2\,\alp{1}{0}\,F_H^+(\vec{u})F_H^-(\p{0})\,\hvec{u}\cdot\hvec{p}_0
\end{pmatrix} \cdot\, [\op{D}_{10}^{+}(\p{0})]^{-1} \, ,
\label{Q+-}
\end{align}
\begin{align}
&\op{Q}^{--}(\vec{u}|\p{0})
=\frac{(\ep_1-\ep_0)}{\alp{0}{0}}\,[\op{D}_{10}^{+}(\vec{u})]^{-1}\,\cdot \no
\\ & \begin{pmatrix}
\ep_0\,\alp{1}{0}\,||\vec{u}||||\p{0}||F_V^-(\vec{u})F_V^-(\p{0}) &
-\ep_0^{\frac{1}{2}}\,K_0\,\alu{1}{}\,\alp{1}{0}\,F_V^+(\vec{u})F_H^-(\p{0})\,
(\hvec{u}\times\hvec{p}_0)_z\\
-\ep_1\,\alu{1}{}\,\alpha_0^2(\p{0})\,F_V^+(\vec{u})F_V^+(\p{0})\,\hvec{u}\cdot
\hvec{p}_0 & \\
& \\
-\ep_0^{-\frac{1}{2}}\,\ep_1\,K_0\,\alpha_0^2(\p{0})\,F_H^-(\vec{u})F_V^+(\p{0})
\, (\hvec{u}\times\hvec{p}_0)_z
&K_0^2\,\alp{1}{0}\,F_H^-(\vec{u})F_H^-(\p{0})\,\hvec{u}\cdot\hvec{p}_0
\end{pmatrix}\cdot \,
[\op{D}_{10}^{+}(\p{0})]^{-1}\, ,
\label{Q--}
\end{align}
where
\begin{equation}
\begin{pmatrix}
F_V^{\pm}(\p{0}) & 0 \\
0 & F_H^{\pm}(\p{0})
\end{pmatrix}
=\left(\op{1}\pm\op{V}^{H\,21}(\p{0})\right)\left[\op{1}+\op{V}^{10}(\p{0})
\cdot \op{V}^{H\,21}(\p{0})\right]^{-1}
\label{Fcoef}
\end{equation}
and 
\begin{align}
\op{P}^{+}(\vec{u}|\p{1})&=(\ep_1-\ep_0)\,[\op{D}_{10}^{+}(\vec{u})]^{-1}\cdot
\begin{pmatrix}
||\vec{u}||||\p{1}||F_V^{+}(\vec{u}) &
-\ep_0^{\frac{1}{2}}\,K_0\,\alu{1}{}\,F_V^{-}(\vec{u})\,(\hvec{u}\times
\hvec{p}_1)_z \\
+\alu{1}{}\,\alp{0}{1}\,F_V^{-}(\vec{u})\,
\hvec{u}\cdot\hvec{p}_1 & \\ & \\
\ep_0^{-\frac{1}{2}}\,K_0\,\alp{0}{1}\,F_H^{+}(\vec{u})\,(\hvec{u}\times
\hvec{p}_1)_z
& K_0^2\,F_H^{+}(\vec{u})\,\hvec{u}\cdot\hvec{p}_1
\end{pmatrix}\, ,\label{P+} \\
\op{P}^{-}(\vec{u}|\p{1})&=(\ep_1-\ep_0)\,[\op{D}_{10}^{+}(\vec{u})]^{-1}\cdot
\begin{pmatrix}
||\vec{u}||||\p{}||F_V^{-}(\vec{u}) &
-\ep_0^{\frac{1}{2}}\,K_0\,\alu{1}{}\,F_V^{+}(\vec{u})\,(\hvec{u}\times
\hvec{p}_1)_z \\
+\alu{1}{}\,\alp{0}{1}\,F_V^{+}(\vec{u})\,
\hvec{u}\cdot\hvec{p}_1 & \\ & \\
\ep_0^{-\frac{1}{2}}\,K_0\,\alp{0}{1}\,F_H^{-}(\vec{u})\,(\hvec{u}\times
\hvec{p}_1)_z
& K_0^2\,F_H^-(\vec{u})\,\hvec{u}\cdot\hvec{p}_1
\end{pmatrix}\, .
\label{P-}
\end{align}  
The first-order term was recently derived by Fuks {\it et al}~\cite{Fuks}. 
They have noticed that for this order, the matrix
differs from the one obtained for a surface separating two
semi-infinite media by only the factors $F^{\pm}$. Likewise, for higher
order, we see that \Tref{Q++}{Q-+} differ from \Eref{Q+} by only
$F^{\pm}$, similarly for  \Tref{Q+-}{Q--} with respect to \Eref{Q-}, and 
\Tref{P+}{P-} with respect to \Eref{P}. 
So when the thickness $H$ becomes infinite,
and the absorption $Im(\ep_1) \neq 0$, or if $\ep_1=\ep_2$, we have
$\op{V}^{H\,21}=0$, thus $F^{\pm}=1$, and in that case we recover the matrix
\sref{Q+}{P} for a rough surface between two semi-infinite media.
%%%%%%%%%%%%%%%%%%%%%%%%%%%%%%%%%%%%%%%%%%%%%%%%%%%%%%%%%%%%%%%%%%%%%%%%%%%%%%%%
\section{The Mueller matrix cross section and the surface statistic}
%%%%%%%%%%%%%%%%%%%%%%%%%%%%%%%%%%%%%%%%%%%%%%%%%%%%%%%%%%%%%%%%%%%%%%%%%%%%%%%%
\label{Muller}
When we consider an observation point in the far field limit, the saddle-point
method gives an asymptotic form for the scattered field
$\vec{E}^s\equiv \vec{E}^{0+}$ obtained from \Eref{DefEp} :

\begin{equation}
\vec{E}^s(\vec{x},z)=\frac{\exp(\rmi K_0||\vec{r}||)}{||\vec{r}||} \,
\op{f}(\p{}|\p{0})\cdot\vec{E}^i(\p{0})\, , 
\end{equation}
with
\begin{align}
\op{f}(\p{}|\p{0})&\equiv \frac{K_0 \cos \theta}{2\pi \rmi} \,
\Rp{}{0} \, ,\\
\vec{p}&=K_0 \, \frac{\vec{x}}{||\vec{r}||} \, ,
\end{align}
where $\theta$ is the angle between $\ez$ and the scattering direction
(see \Fref{Vectonde}).
In order to describe the incident and the scattered waves, we
introduce the modified Stokes parameters:
\begin{equation}
\vec{I}^s(\p{})\equiv\left(\begin{array}{c}
|E_V^s(\p{})|^2 \\ |E_H^s(\p{})|^2\\
2Re(E^s_V(\p{})E^s_H(\p{}))\\
2Im(E^s_V(\p{})E^s_H(\p{}))\end{array} \right)
,\quad
\vec{I}^i(\p{0})\equiv\left( \begin{array}{c}
|E_V^i(\p{0})|^2 \\ |E_H^i(\p{0})|^2\\
2Re(E^i_V(\p{0})E^i_H(\p{0}))\\
2Im(E^i_V(\p{0})E^i_H(\p{0}))\end{array} \right) \, .
\end{equation}
The analog of the scattering matrix for these parameters is the
Mueller matrix, defined~\cite{Ishi} by:
\begin{equation}
\vec{I}^s(\p{}) \equiv \frac{1}{||\vec{r}||^2} \op{M}(\p{}|\p{0}) \cdot
\vec{I}^i(\p{0})\, ,
\end{equation}
which can be expressed as a  function~\cite{Ishi} of $\op{f}(\p{}|\p{0})$.
To maintain a matrix formulation in the following calculations, we
introduce a new product between two-dimensional matrices with the definition :
\begin{align}
\op{f}\odot \op{g} &\equiv  \begin{pmatrix} f_{VV} & f_{VH} \\f_{HV}
& f_{HH} \end{pmatrix} \odot \begin{pmatrix} g_{VV} & g_{VH}
\\g_{HV} & g_{HH} \end{pmatrix} \\
&= \left(\begin{array}{cccc} 
f_{VV}g^*_{VV} & f_{VH} g_{VH}^* & Re(f_{VV}g_{VH}^*) & -Im(f_{VV}g_{VH}^*)\\
f_{HV}g^*_{HV} & f_{HH} g_{HH}^* & Re(f_{HV}g_{HH}^*) & -Im(f_{HV}g_{VH}^*)\\
2Re(f_{VV}g^*_{HV}) & 2Re(f_{VH} g_{HH}^*) & Re(f_{VV} g_{VV}^*+f_{HV} g_{VH}^*)
& -Im(f_{VV}g_{HH}^*-f_{VH}g_{HV}^*)\\
2Im(f_{VV}g^*_{HV}) & 2Im(f_{VH} g_{HH}^*) & Im(f_{VV} g_{VV}^*+f_{HV} 
g_{VH}^*)& 
Re(f_{VV}g_{HH}^*-f_{VH}g_{HV}^*)\end{array} \right) \, . \nonumber
\end{align}
This product allows to express the matrix $\op{M}$ as :
\begin{align}
\op{M}(\p{}|\p{0})& =\op{f}(\p{}|\p{0})\odot\op{f}(\p{}|\p{0}) \, ,\\
&=\frac{K^2_0 \cos^2
\theta}{(2\pi)^2}\,\op{R}(\p{}|\p{0})\odot\op{R}(\p{}|\p{0})\, .
\end{align}
Following Ishimaru {\it et al.}~\cite{Ishi2}, we define 
the Mueller matrix cross section   
per unit area $\op{\sigma}=(\sigma_{ij})$ :
\begin{equation}
\op{\sigma}\equiv\frac{4\pi}{A}\, \op{M}\, ,
\end{equation}
and, the bistatic Mueller matrix~\cite{Kong} $\op{\ga}=(\ga_{ij})$:
\begin{equation}
\op{\gamma}\equiv \frac{1}{A \cos \theta_0}\, \op{M}\, .
\end{equation}

These matrix are the generalization of the classical coefficients. In
fact, if we assume, for example, that the incident wave is vertically
polarized we have:
\begin{align}
\frac{1}{A\,\cos
\theta_0}|E_V^s(\p{})|^2&=\frac{1}{||r||^2}\ga_{11}(\p{}|\p{0})|
E_V^i(\p{})|^2 \, , \\
\frac{1}{A\,\cos
\theta_0}|E_H^s(\p{})|^2&=\frac{1}{||r||^2}\ga_{21}(\p{}|\p{0})|
E_V^i(\p{})|^2 \, .
\end{align}
Thus $\gamma_{11}$ and $\gamma_{21}$, are respectively the classical
bistatic coefficients  $\gamma_{VV}$ and $\gamma_{HV}$. 
We can also define the cross section and the bistatic coefficients for an 
incident circular polarization. 
As an example, taking the incident wave right circularly polarized, we have 
\begin{equation}
\vec{I}^i(\p{0})=\frac{1}{2}\,(\begin{array}{cccc}1& 1&
0& -2\end{array})^t \, ,
\end{equation}
now, if we put a
right-hand side polarizer at the receiver:
\begin{align}
\frac{1}{4}\left(\begin{array}{cccc}
1 & 1 & 0 & 1 \\
1 & 1 & 0 & 1 \\
0 & 0 & 0 & 0 \\
2 & 2 & 0 & 2 
\end{array}\right)\, ,
\end{align}
the right to right bistatic coefficient $\ga_{rr}$ is:
\begin{equation}
\ga_{rr}=\frac{1}{4}\left(\ga_{11}+\ga_{12}+2\,\ga_{14}+\ga_{21}+\ga_{22}+2\,
\ga_{24}+\ga_{41}+\ga_{42}+2\,\ga_{44}\right)\, ,
\end{equation}
where $\ga_{ij}$ are coefficients of the matrix
$\op{\ga}=(\ga_{ij})$.

In a similar way we obtain the right to left bistatic coefficient:
\begin{equation}
\ga_{lr}=\frac{1}{4}\left(\ga_{11}+\ga_{12}+2\,\ga_{14}+\ga_{21}+\ga_{22}+2\,
\ga_{24}-\ga_{41}-\ga_{42}-2\,\ga_{44}\right)\, .
\end{equation}

Up to now, we have made no hypothesis on the nature of the
rough surface.
Let us introduce the statistical caracteristics  of the function $\hx$. 
We suppose that it is a stationary, isotropic Gaussian random 
process defined by the moments 
\begin{align}
<\hx>&=0,\\
<h(\x)\,h(\vec{x'})>&=W(\vec{x}-\vec{x'})\, ,
\end{align}
where the angle brackets denote an average over the ensemble of
realizations of the function $\hx$.
In this work we will use a Gaussian form for the surface-height correlation 
function $W(\x)$:
\begin{equation}
W(\x)=\sigma^2\,\e(-\x^2/l^2)\, ,
\end{equation}
where $\sigma$ is the rms height of the surface, and $l$ is the
transverse correlation length. In momentum space we have:
\begin{align}
<h(\p{})>&=0,\\
<h(\p{})\,h(\vec{p}')> &=(2\pi)^2\,\delta(\p{}+\vec{p}')\,W(\p{})\, ,
\end{align}
with 
\begin{align}
W(\p{})&\equiv\int\,\rmd^2\x\,W(\x)\,\e(-i\p{}\cdot\x)\\
&=\pi\,\sigma^2\,l^2\,\e(-\p{}^2\,l^2/4)\, .
\end{align}
We are now able to define the bistatic coherent matrix
\begin{align}
\op{\ga}^{coh}&\equiv \frac{1}{A\,\cos
\theta_0}\,<\op{f}(\p{}|\p{0})>\odot<\op{f}(\p{}|\p{0})>
=\frac{K^2_0 \cos^2 \theta}{A\,(2\pi)^2\,\cos
\theta_0}\,<\op{R}(\p{}|\p{0})>\odot<\op{R}(\p{}|\p{0})>\, ,
\end{align}
and the incoherent bistatic matrix 

\begin{align}
\op{\ga}^{incoh}(\p{}|\p{0})&\equiv  \frac{1}{A\,\cos
\theta_0}\,[<\op{f}(\p{}|\p{0})\odot\op{f}(\p{}|\p{0})>-<\op{f}(\p{}|\p{0})>
\odot<\op{f}(\p{}|\p{0})>] \, ,\no \\
&=\frac{K^2_0 \cos^2 \theta}{A\,(2\pi)^2\,
\cos \theta_0}\left[<\op{R}(\p{}|\p{0})\odot\op{R}(\p{}|
\p{0})>-<\op{R}(\p{}|\p{0})>\odot<\op{R}(\p{}|\p{0})>\right]\, .
\label{incoh}
\end{align}

From \Eref{Dev1}, and the property of the Gaussian random process, we obtain
\begin{align}
\op{\ga}^{coh}(\p{}|\p{0})
&=\frac{K^2_0 \cos^2 \theta}{\cos
\theta_0}\,\delta(\p{}-\p{0})\,\op{R}^{coh}(\p{0})\,\odot\op{R}^{coh}(\p{0}) \\
\op{R}^{coh}(\p{0})&\equiv \op{X}^{(0)}(\p{0})+K_0\,\cos
\theta_0\,\intp{}\,\op{X}^{(2)}(\p{0}|\p{1}|\p{0})\,W(\p{1}-\p{0})+\cdots \, ,
\end{align}
where $\op{R}^{coh}(\p{0})$ is a diagonal matrix describing the
reflection coefficients of the coherent waves.
For the incoherent part we have :
\begin{align}
\op{\ga}^{incoh}(\p{}|\p{0})
&=\frac{K^4_0 \cos^2 \theta\,\cos
\theta_0}{(2\pi)^2}\,[\op{I}^{(1-1)}(\p{}|\p{0})+\op{I}^{(2-2)}(\p{}|\p{0})+
\op{I}^{(3-1)}(\p{}|\p{0})]\, ,\label{incoh2}
\end{align}
where
\begin{align}
\op{I}^{(1-1)}(\p{}|\p{0})&\equiv W(\p{}-\p{0})\,\op{X}^{(1)}(\p{}|\p{0})
\odot\op{X}^{(1)}
(\p{}|\p{0})\label{ich1}\\ 
\op{I}^{(2-2)}(\p{}|\p{0})&\equiv \intp{1}\,W(\p{}-\p{1})W(\p{1}-\p{0})\,
\op{X}^{(2)}(\p{}|\p{1}|\p{0})\odot
\left[\op{X}^{(2)}(\p{}|\p{1}|\p{0})+\op{X}^{(2)}(\p{}|\p{}+\p{0}-\p{1}|
\p{0})\right]\label{ich2}\\
\op{I}^{(3-1)}(\p{}|\p{0})&\equiv W(\p{}-\p{0})\left[\op{X}^{(1)}(\p{}|\p{0})
\odot\op{X}^{(3)}(\p{}|\p{0})
+\op{X}^{(3)}(\p{}|\p{0})\odot
\op{X}^{(1)}(\p{}|\p{0})\right]\label{ich3}\, ,
\end{align}
with 
\begin{align}
\op{X}^{(3)}(\p{}|\p{0})\equiv \intp{1}\,&\left[W(\p{1}-\p{0})\,
\op{X}^{(3)}(\p{}|\p{0}|\p{1}|\p{0}) \right. \no \\
&\left.+W(\p{}-\p{1})\,\left(\op{X}^{(3)}(\p{}|\p{1}|\p{0}-\p{}+
\p{1}|\p{0})+\op{X}^{(3)}(\p{}|\p{1}|\p{}|\p{0})\right)\right]\, .
\end{align}
%%%%%%%%%%%%%%%%%%%%%%%%%%%%%%%%%%%%%%%%%%%%%%%%%%%%%%%%%%%%%%%%%%%%%%%%%%%%%%%
\section{Applications}
%%%%%%%%%%%%%%%%%%%%%%%%%%%%%%%%%%%%%%%%%%%%%%%%%%%%%%%%%%%%%%%%%%%%%%%%%%%%%%%
\label{Numer}
In the previous sections we have developed a method to compute the scattering
matrices for a rough surface between two media, and for a thin film which 
includes one rough surface. In this section, we will evaluate numerically
the incoherent bistatic coefficients given by \sref{incoh2}{ich3} for different
values of the parameters which characterize the configurations. 
In all numerical simulations the media 0 will be the vacuum ($\ep_0=1$).

\vskip 0.2cm
%%%%%%
\noindent \underline{A rough surface separating to different media}
\vskip 0.2cm

\begin{figure}[ht]
% figure 8
\includegraphics{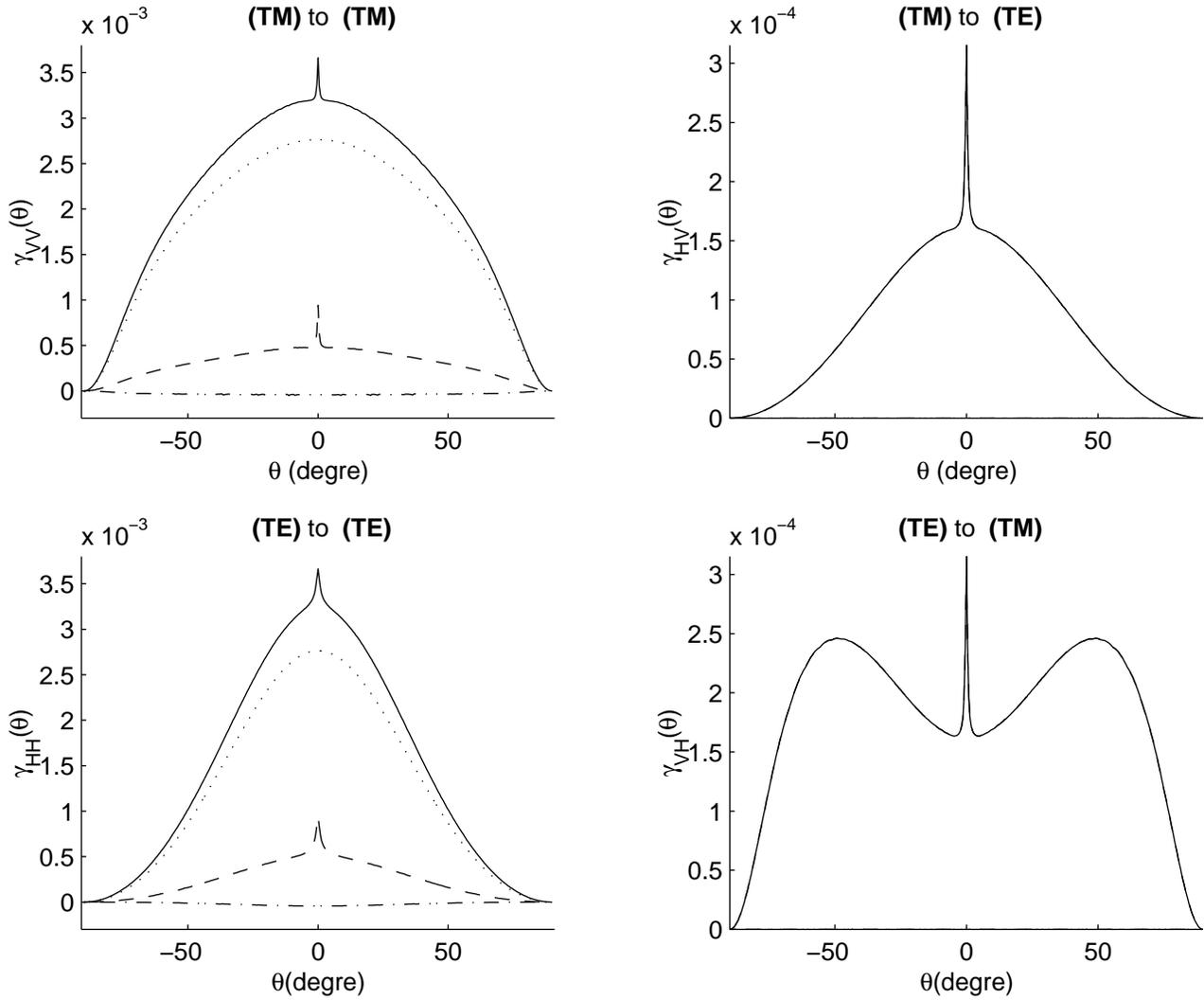}
\caption{The bistatic coefficients for an horizontal $(TE)$ and a vertical 
$(TM)$ polarized incident light of wavelength $\lambda=457.9\,nm$ 
($\theta_0=0\degre$ and $\phi=\phi_0=0\degre$), on a
two-dimensional randomly rough silver surface, characterized by the
parameters $\sigma=5 \,nm$, $l=100\,nm$. $\epsilon_1=-7.5+i0.24$.  
For each figure are plotted : the total incoherent
scattering $\op{\ga}^{incoh}$ (solid curve), the first order given by
$\op{I}^{(1-1)}$(dotted line), 
the second order $\op{I}^{(2-2)}$ (dashed line), and the
third order $\op{I}^{(3-1)}$ (dash-dotted line).}
\label{simpleLin}
\end{figure}

We consider that a polarized light of wavelength $\lambda=457.9\,nm$ 
is normally incident ($\theta_0=0\degre,\phi_0=0\degre$) on a two-dimensional
rough silver surface (see \Fref{surfaceSimple}) characterized by the 
roughness parameters $\sigma=5\, nm$, $l=100\, nm$, $\epsilon=-7.5+i0.24$. 
As a matter of comparison we have chosen the same parameters
used for the scattering by a one-dimensional rough surface~\cite{Mendez}.
The perturbative development is given by \Tref{ordre1}{P}.
In \Fref{simpleLin}, we present the results for an incident wave linearly
polarized, the scattered field being observed in the incident plane
($\phi=0\degre$).
The single scattering contribution associated with the term 
$\op{I}^{(1-1)}$ is plotted as a dotted line, the double-scattering
contribution $\op{I}^{(2-2)}$ as a dashed line, the scattering term
$\op{I}^{(3-1)}$ as a dash-dotted line, and the sum of all these terms
$\op{\ga}^{incoh}$ by the solid curve.

\begin{figure}[ht]
\includegraphics[height=8cm,width=14cm]{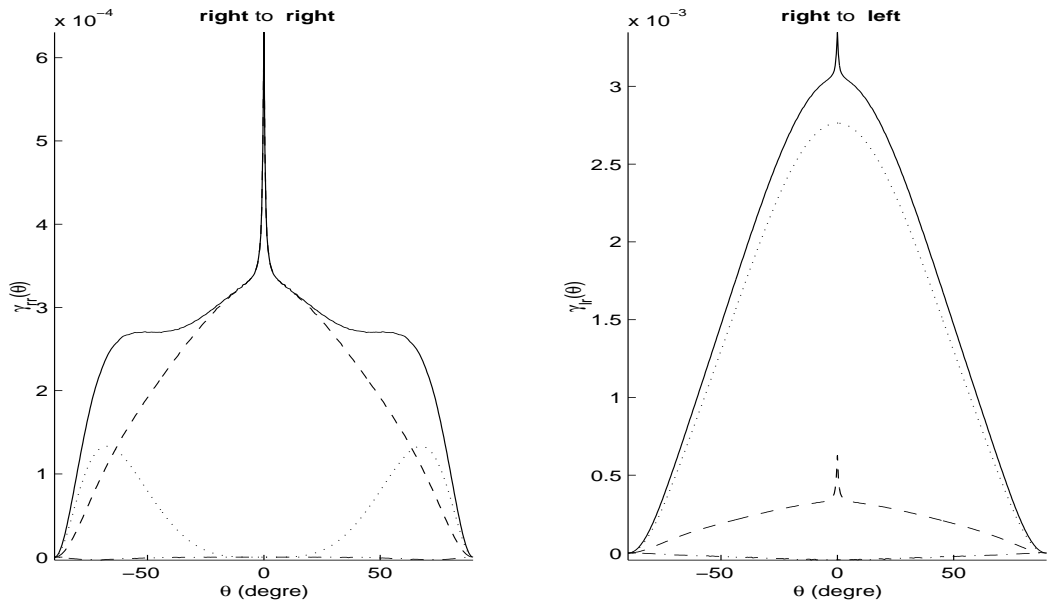}
\caption{The same configuration as \Fref{simpleLin}, 
but with a right incident cicularly polarized wave, and a right to right
(or left to left), right to left (or left to right) observed polarizations}
\label{simpleCir}
\end{figure}

We observed an enhancement of the backscattering which corresponds to 
the physical process in which the incident light excites a surface
electromagnetic wave. In fact, the surface polariton propagates along
the rough surface, then it is scattered into a volume wave due to the
roughness, at the same time, a reverse partner exists with a path
travelling in the opposite direction. These two paths can interfere
constructively near the backscattering direction to produce a
peak~\cite{Mara1,Mara2,Mara3}. However, in one dimension~\cite{Mendez}, 
this peak can only be observed for a $(TM)$ polarized incident wave because a
surface polariton only exists for this polarization. 
In two dimensions, the surface wave also exists for a $(TM)$ polarization, 
but in fact a depolarization occurs so that a $(TE)$ incident wave 
can excite a $(TM)$ surface wave, 
and this surface wave can be scattered into volume 
wave with both polarizations as can be seen in \Fref{simpleLin}.
Now, when the incident wave is circularly polarized, we see in
\Fref{simpleCir}, that the enhanced backscattering takes also place. 
We have not displayed the left to left, and left to right
polarizations because the media are not optically active,
as a consequence, the results are the same either the incident wave 
is right or left polarized.

In the expression \eref{incoh2}, the peak is produced by the term 
$\op{I}^{(2-2)}$. We see that the term $\op{P}(\vec{u}|\p{1})
\cdot\op{Q}^{+}(\p{1}|\p{0})$ in 
$\op{X}^{(2)}_{s\, \ep_0,\ep_1}(\p{}|\p{1}|\p{0})$ 
contains a factor of the form (see \Eref{D10}):
\begin{equation}
[D^{+}_{10\,VV}(\p{1})]^{-1}=\frac{1}{\ep_1\alp{0}{1}+\ep_0\alp{1}{1}},
\end{equation}
which is close to zero excepted when $\p{1}$ is near the resonance mode $\p{r}$
of the polariton, which is given by the roots
$D^{+}_{10\,V}(\p{r})=0$.
When we observe the field scattered far away from the backscattering
direction ($\p{}+\p{0}\not = 0$), the terms 
$\op{X}^{(2)}_{s\, \ep_0,\ep_1}(\p{}|\p{1}|\p{0})$ and 
$\op{X}^{(2)}_{s\, \ep_0,\ep_1}(\p{}|\p{}+\p{0}-\p{1}|
\p{0})$ contening $D^{+}_{10\,V}$ are non zero when 
$\p{1}\approx\p{r}$, and $\p{}+\p{0}-\p{1}\approx\p{r}$ respectively. 
Since these domains are disjoint,
the product $\odot$ of these two terms is approximatively zero.
Conversely, when we are near the backscattering direction
($\p{}+\p{0}\approx 0$), the terms inside brackets are almost equal and
produce the enhancement factor.
This enhancement factor is not equal to 2 because the matrices
$\op{Q}^{+}(\p{}|\p{0})$ and $\op{Q}^{-}(\p{}|\p{0})$ in
$\op{X}^{(2)}_{s\,\ep_0,\ep_1}$ 
do not contain the term $[D^{+}_{10\,V}(\p{1})]^{-1}$,
so they produce a significative contribution whatever the scattering
angle is. 
In order to isolate more precisely the terms producing an enhanced
backscattering, a better approach is to work with the formalism of
\citeo{Celli} derived from quantum mechanical scattering theory, such an
approach is used for instance in \citeo{Mara1,Sanchez}. 
If the decomposition of each step of the mutiple scattering process is 
clearly put in evidence, however, it offers the disadvantage to 
produce a more heavier perturbative development as it can be seen when
comparing \eref{Dev1} and \sref{Q+}{P} with (15-19,A-1) of \citeo{McGurn}.  

\vskip 0.2cm
%%%%%
\noindent \underline{A film with a rough surface on the upper side}
\vskip 0.2cm

\begin{figure}[ht]
\includegraphics[width=16cm,height=14cm]{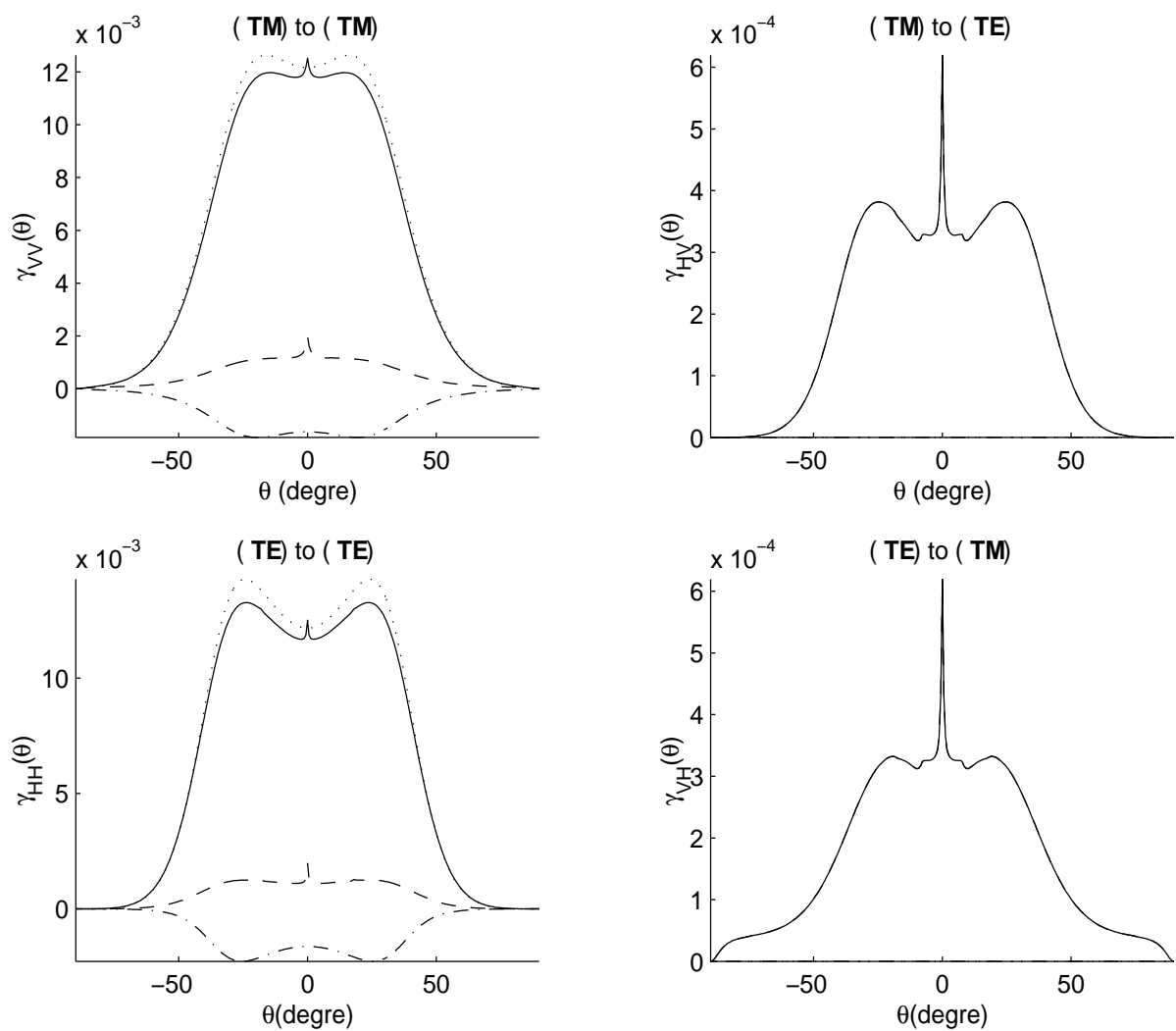}
\caption{The bistatic coefficients for an horizontal $(TE)$ and a vertical 
$(TM)$ polarized incident light of wavelength $\lambda=632.8\,nm$, on a slab
with an upper two-dimensional randomly rough surface, characterized by the
parameters $\sigma=15 \,nm$, $l=100\,nm$, $\epsilon_1=2.6896+i0.0075$, 
thickness $H=500nm$, deposited  on an infinite conducting plane 
($\ep_2=-\infty$). 
The scattered field is observed in the incident plane. 
For each figure are plotted : the total incoherent
scattering $\op{\ga}^{incoh}$ (solid curve), the first order given by
$\op{I}^{(1-1)}$ (dotted line), 
the second order $\op{I}^{(2-2)}$ (dashed line), and the
third order $\op{I}^{(3-1)}$ (dash-dotted line).}
\label{SlabSup}
\end{figure}

\begin{figure}[ht]
\includegraphics[width=16cm,height=14cm]{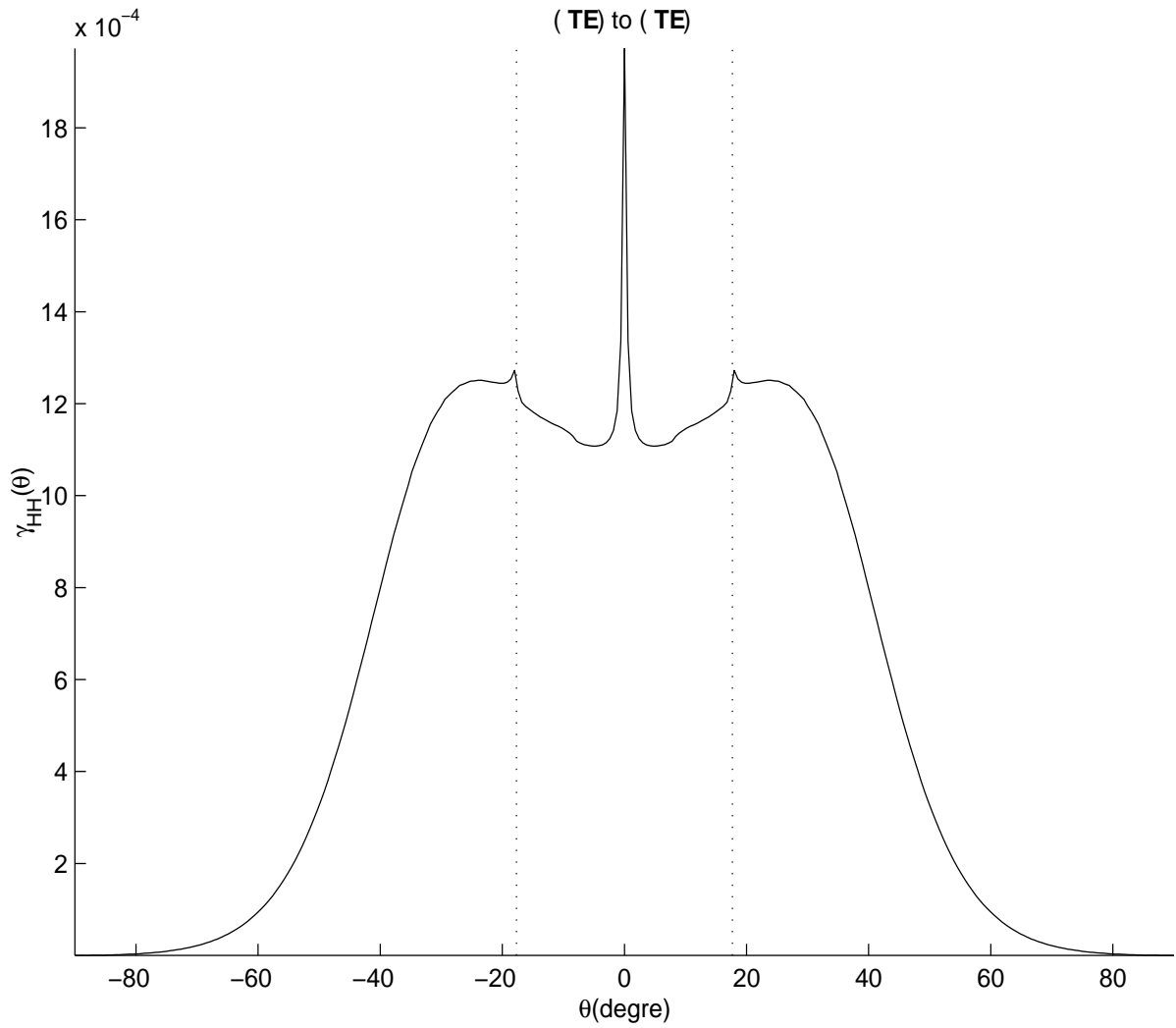}
\caption{Details of second order $(TE)$ to $(TE)$ contribution to the
  scattering shown in \Fref{SlabSup}. We see two satellite peaks at 
  the angle $\theta^{12}_{\pm}(TE)=\pm\,17.7\degre$, the dotted-lines
  mark the peaks angle position.}
\label{SlabSup2}
\end{figure}

\begin{figure}[ht]
\includegraphics[width=16cm,height=14cm]{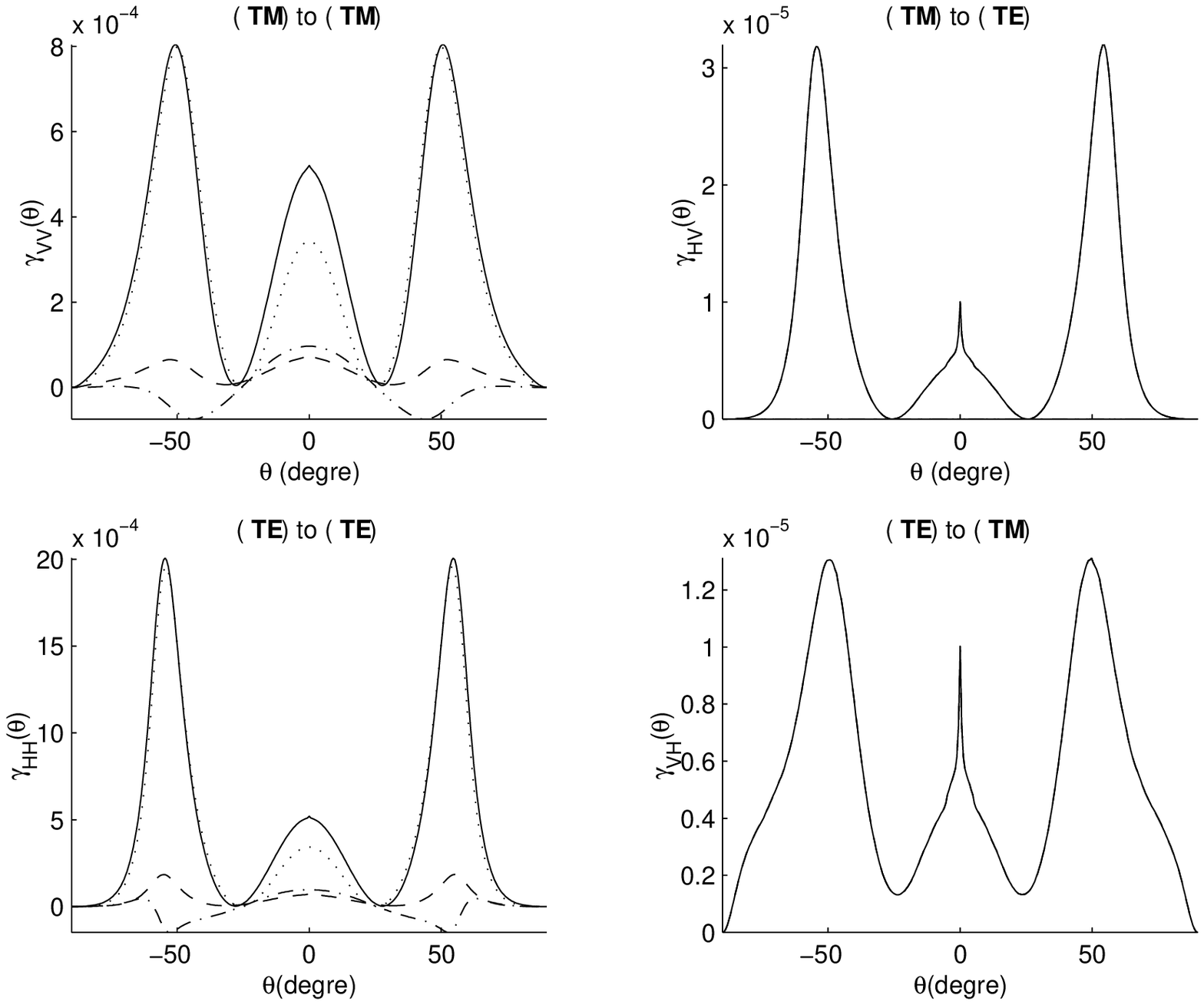}
\caption{Effect of the slab thickness, $H=1000nm$, on the configuration
shown in \Fref{SlabSup}.}
\label{SlabSup2H}
\end{figure}

We consider a dielectric film (see \Fref{slabU}) of mean thickness
$H=500\,nm$, dielectric constant $\epsilon_1=2.6896+0.0075i$, deposited 
on a planar perfectly conducting substrate ($\ep_2=-\infty$) and
illuminated by a linearly polarized light of wavelength
$\lambda=632.8\,nm$ normaly incident ($\phi_0=0\degre$, $\theta_0=0\degre$).
The two-dimensional upper rough surface is characterized by the
parameters $\sigma=15\,nm$ and $l=100\,nm$. The scattering diagrams are 
shown in \Fref{SlabSup} with the same curve labelling as before.
The perturbative development being given by \Tref{ordre1slab}{ordre3slab}
and \Tref{Q++}{P-}. Since we have chosen an infinite conducting plane
($\ep_2=-\infty$) the coefficients $F^{\pm}$ (\Eref{Fcoef}) 
have the following form :
\begin{align}
F_V^{\pm}(\p{0})&=\frac{1\pm \e(2\,\rmi\,\alp{0}{0}\,H)}
{(\ep_1\,\alp{0}{0}+\ep_0\,\alp{1}{0})+
(\ep_1\,\alp{0}{0}-\ep_0\,\alp{1}{0})\,\e(2\,\rmi\,\alp{0}{0}\,H)} 
\label{FVs}\, , \\
F_H^{\pm}(\p{0})&=\frac{1\mp \e(2\,\rmi\,\alp{0}{0}\,H)}
{(\alp{0}{0}+\alp{1}{0})+(\alp{0}{0}-\alp{1}{0})\,\e(2\,\rmi\,\alp{0}{0}\,H)}\, . 
\label{FHs}
\end{align}
The parameters are the same as those used in \citeo{Sanchez} for a 
one-dimensional dielectric film where a $(TE)$ polarized wave is incident.
The thickness was chosen in way that the slab supports only two guided
wave modes, $p_{TE}^1=1.5466\,K_0$, and $p_{TE}^2=1.2423\,K_0$ for the
$(TE)$ polarization.
These modes are resonance modes, they verify
$[F_H^{\pm}]^{-1}(p_{TE}^{1,2})=0$.
For the $(TM)$ case, we have three modes given by the roots
$[F_V^{\pm}]^{-1}(p_{TM})=0$,
which are : $p_{TM}^1=1.6126\,K_0$, $p_{TM}^2=1.3823\,K_0$ and
$p_{TM}^3=1.0030\,K_0$.
As described in \citeo{Mara,Sanchez,Ogura}, these guided modes can
produce a classical enhanced backscattering with satellite peaks
symmetrically positioned. 
The satellite peaks angles are given by the equation :
\begin{equation}
\sin\theta^{nm}_{\pm}=-\sin\theta_0\pm\frac{1}{K_0}\,[p^n-p^m]\, ,
\label{eqpic}
\end{equation}
where $p^n$, $p^m$ describe one of the guided mode, when $n=m$, we
recover the classical enhanced backscattering.
We can give an explanation of this formula as in the previous case.
In the expression \eref{ordre2slab}, the term producing the peaks comes
from 
\begin{equation}
-2\,\op{P}(\vec{u}|\p{1})\cdot\op{Q}^+(\p{1}|\p{0})
\end{equation}  where
$\op{Q}^+(\p{1}|\p{0})$ contains the factors $F^{\pm}(\p{1})$ having
 resonances for the slab guided mode. The product
\begin{equation}
\op{X}_d^{(2)}(\p{}|\p{1}|\p{0})\odot\op{X}_d^{(2)}(\p{}|\p{}+\p{0}-\p{1}|\p{0})
\end{equation}
in \Eref{ich2} has a significant contribution only when $\p{1}$ and $\p{}+\p{0}-\p{1}$
are near resonance modes. As there are several resonances, we can have 
$\p{1}\approx\p{n}$ and $\p{}+\p{0}-\p{1}\approx\p{m}$ with $n\neq m$ where $\p{n}$ 
and $\p{m}$ are resonances vector. If $\p{n}=\pm p_{n}\ex$ and
$\p{m}=\mp p_{m}\ex$ (guided modes propagating along the incident
plane  but with opposite directions), we have $(\p{}+\p{0})\cdot\ex\approx\pm(p_n-p_m)$ which
is another way of writing  \Eref{eqpic}.
For the $(TE)$ polarization, since we have only two guided waves, the
satellite peaks can only exist at the angles
$\theta^{12}_{\pm}(TE)=\pm\,17.7\degre$.
Now, for the $(TM)$ polarization we have three possibilities: 
$\theta^{12}_{\pm}(TM)=\pm 13.3\degre$, $\theta^{13}_{\pm}(TM)=\pm 37.6\degre$, 
and $\theta^{23}_{\pm}(TM)=\pm 22.3\degre$.
The satellite peaks are produced by the term $\op{I}^{(2-2)}$, in the case of
$(TM)$ polarization  we do not get any significant contribution to
satellite peaks. However, for the $(TE)$ to $(TE)$ scattering
shown in \Fref{SlabSup2}, we find satellite peaks at the angle
$\theta^{12}_{\pm}(TE)=\pm\,17.7\degre$ positioned along a dotted line.
Now, by doubling the slab thickness, see \Fref{SlabSup2H}, the
satellite peaks disappear for all the polarization, but we see
a new  phenomenon called the Sel\'enyi fringes.~\cite{Ogura,Frange1,Frange2}. 
For a slightly random rough surface, the slab produces fringes similar
to those obtain with a Fabry-Perrot interferometer illuminated by an
extended source. The roughness modulates amplitude fringes but their
localization remains the same as for the interferometer.
We also notice that the enhanced backscattering decreases with the slab
thickness. We can conclude that as in the case of one-dimensional 
rough surface, the satellite peaks appear only when the wave guide
supports few modes for the $(TE)$ polarizations. 
These results differ from those obtained in \citeo{Ogura} where 
no satellite peak appears in their two-dimensional slab, we have checked that
with their parameters values we also find no peak, and we agree with the
results given by the contributions of the first and second order terms.
However, the third order gives a contribution larger than the first one,
such a result casts some doubt on the validity of the (SPM) method in that
case.

\begin{figure}[ht]
\includegraphics[width=16cm,height=14cm]{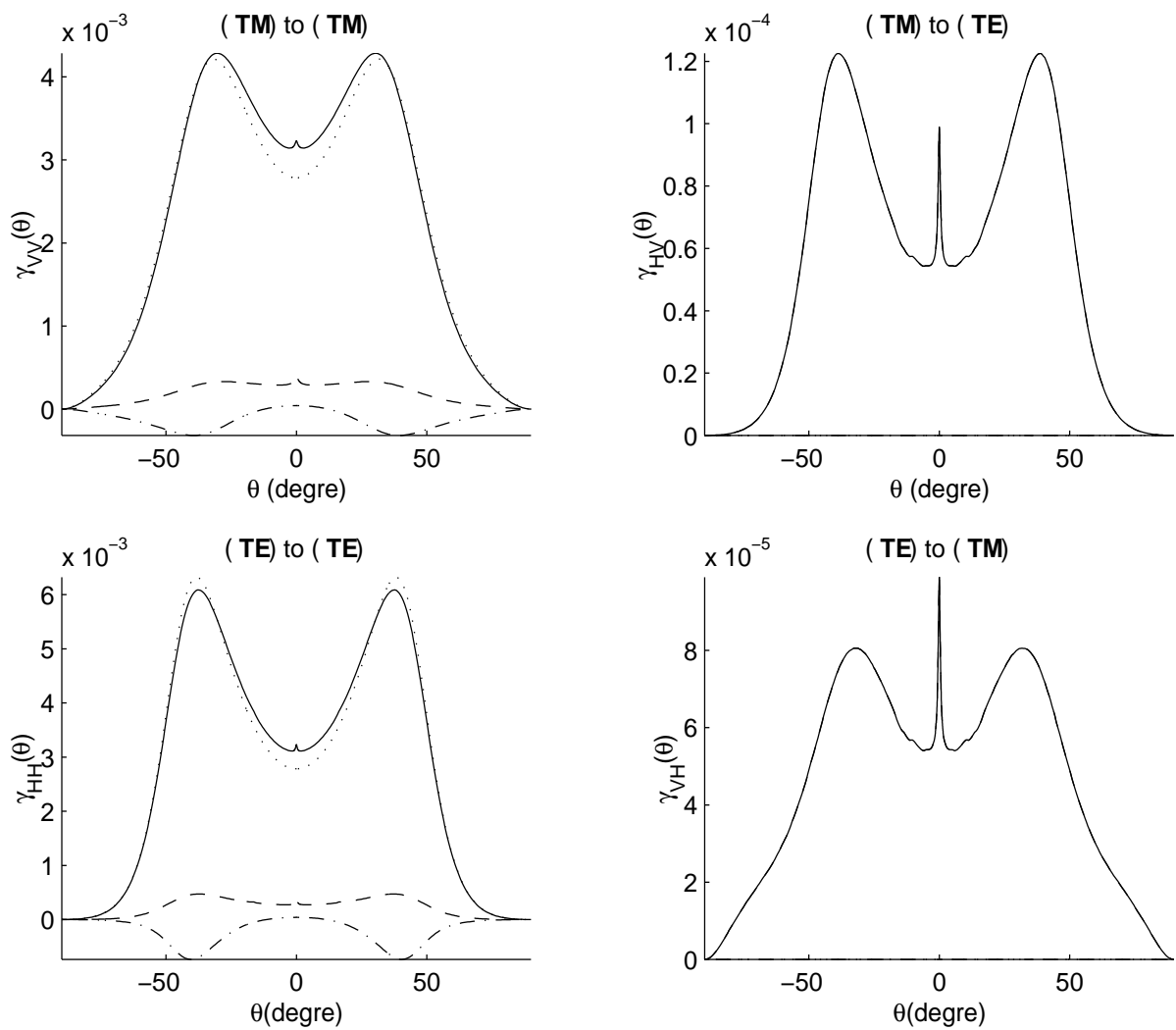}
\caption{The same parameters as in \Fref{SlabSup}, but with a silver 
plane characterized by $\epsilon_2=-18.3+0.55i$.}
\label{SlabSupef}
\end{figure}

However, for the choice of 
parameters presented here no satellite peak has been observed even when
the thickness of the slab is chosen in such a way that only two guided
modes exist for the $(TM)$ polarization (a result not presented here).
This is in agreement with the results of \citeo{Sanchez2} for one
dimensional surface where it is  noticed that the excitation of $(TM)$
modes are more difficult to excite than the $(TE)$ 
modes. In order to enhance this effect they choose a higher
permittivity for the media 1 : $\epsilon_1=5.6644+i0.005$. In this
case satellite peaks are observed for a slab which supports 
three guided modes. We have also done numerical calculations with
these parameters, however we do not observe satellite peaks.
So the transition from one dimensional to two dimensional rough surface
lower the efficiency of the excitation of $(TM)$ modes. 
Next, instead of doubling the slab thickness, we have changed the
infinite conducting plane by a silver plane ($\epsilon_2=-18.3+0.55i$),
we see in \Fref{SlabSupef}, that the enhancement of backscattering is also 
decreased, and that there is no more satellite peak corresponding to $(TE)$ to
$(TE)$ scattering.
This fact has to be compared with the next configuration, where the rough
surface is now between the media 1 and the media 2, see \Fref{slabB}. 

\vskip 0.2cm
%%%%%
\noindent \underline{A film with a rough surface on the bottom side}
\vskip 0.2cm

\begin{figure}[ht]
\includegraphics[width=16cm,height=14cm]{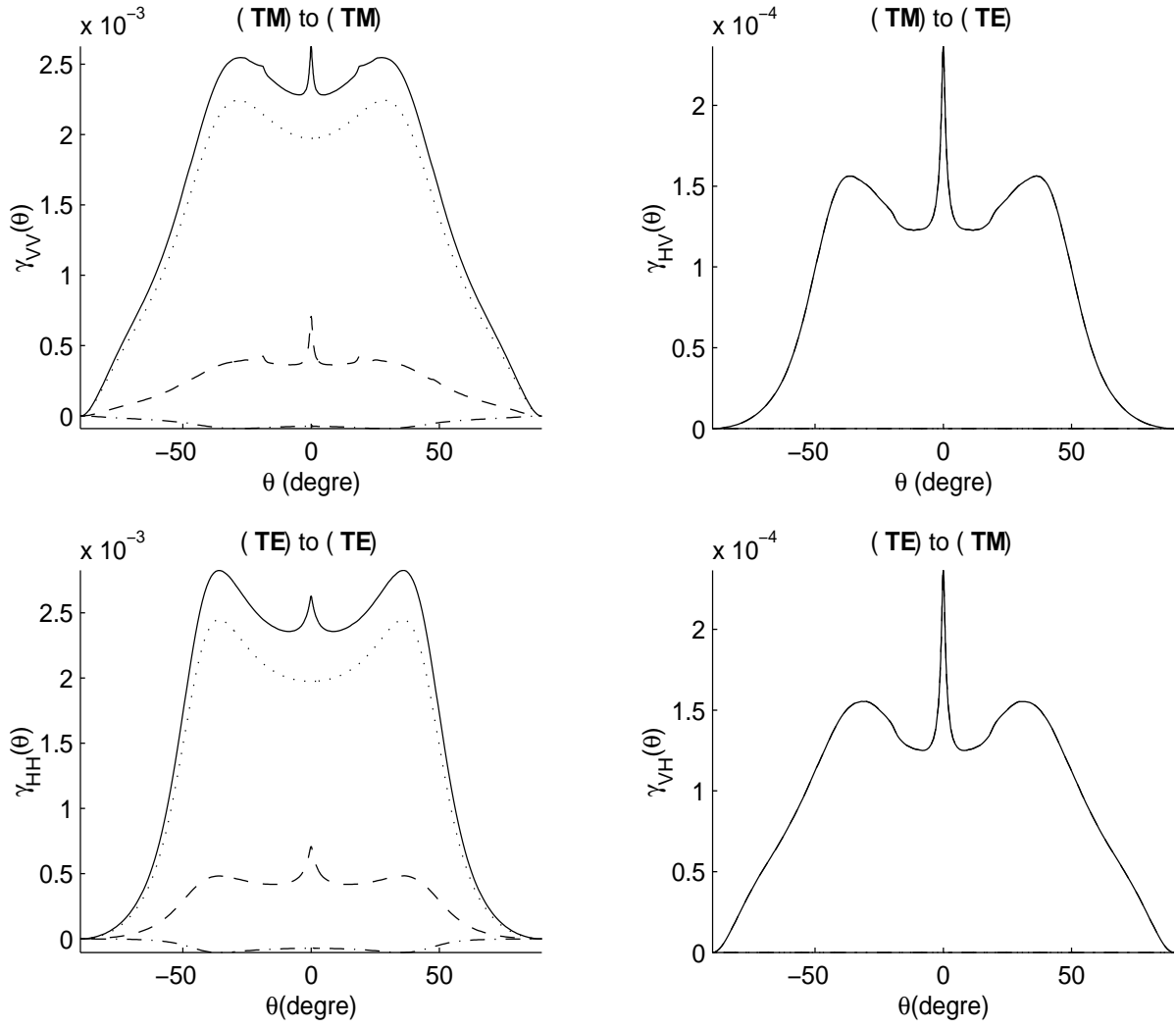}
\caption{The bistatic coefficients for an horizontal $(TE)$ and vertical 
$(TM)$ polarized light of wavelength $\lambda=632.8\,nm$, incident on a film
of permittivity $\epsilon_1=2.6896+i0.0075$, deposited on a 
two-dimensional randomly rough surface, characterized by the
parameters, $\sigma=5 \,nm$, $l=100\,nm$,  $\epsilon_2=-18.3+0.55i$,
thickness $H=500nm$.
The scattered field is observed in the incident plane. 
For each figure are plotted : the total incoherent
scattering $\op{\ga}^{incoh}$ (solid curve), the first order given by
$\op{I}^{(1-1)}$ (dotted line), 
the second order $\op{I}^{(2-2)}$ (dashed line), and the
third order $\op{I}^{(3-1)}$(dash-dotted line).}
\label{SlabInf}
\end{figure}

\begin{figure}[ht]
\includegraphics[width=16cm,height=14cm]{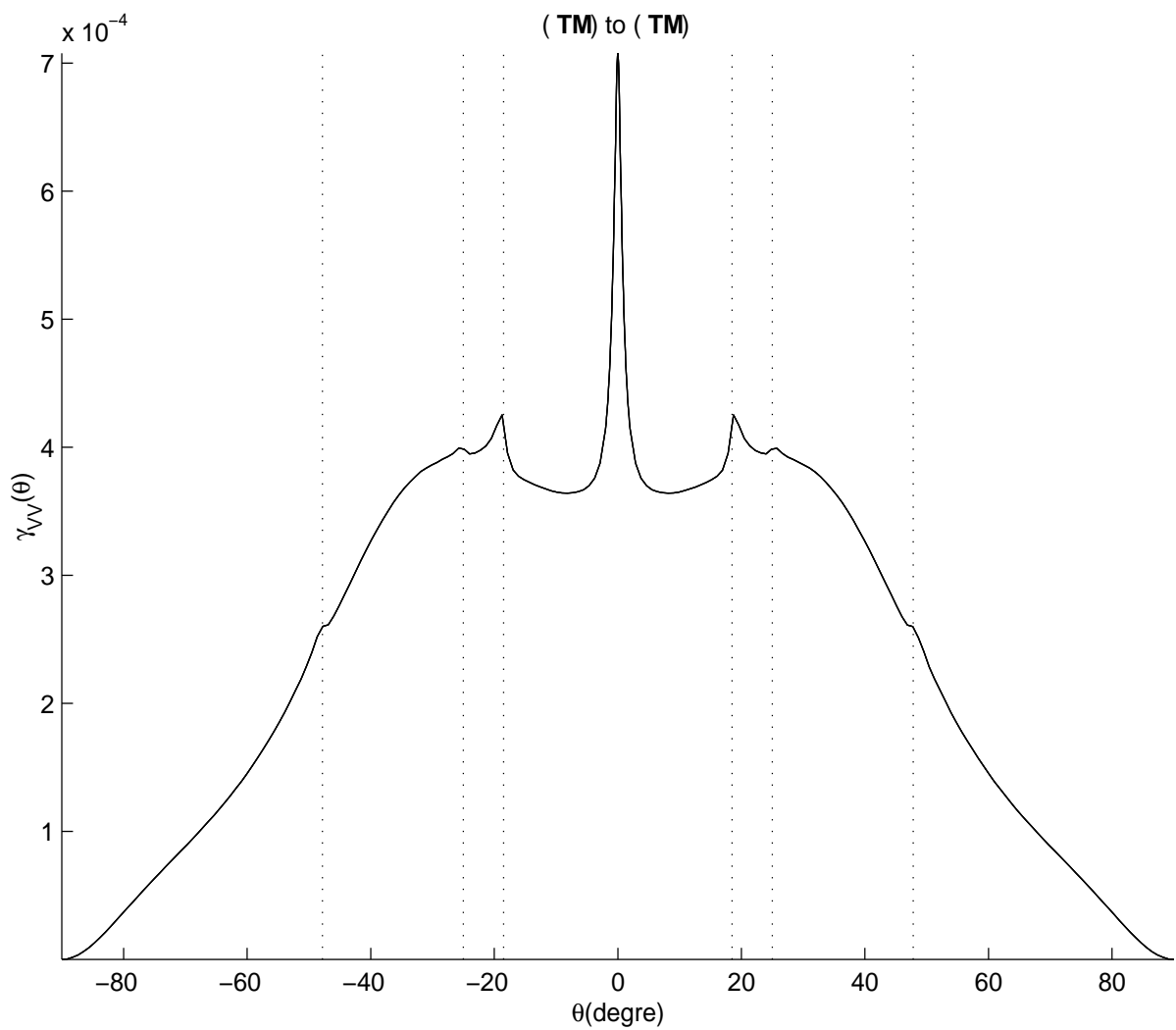}
\caption{Details of the second order $(TM)$ to $(TM)$ contribution to
the scattering shown in \Fref{SlabInf}, dotted lines mark the 
peaks angle position.}
\label{SlabInf2}
\end{figure}

\begin{figure}[ht]
\includegraphics[width=16cm,height=14cm]{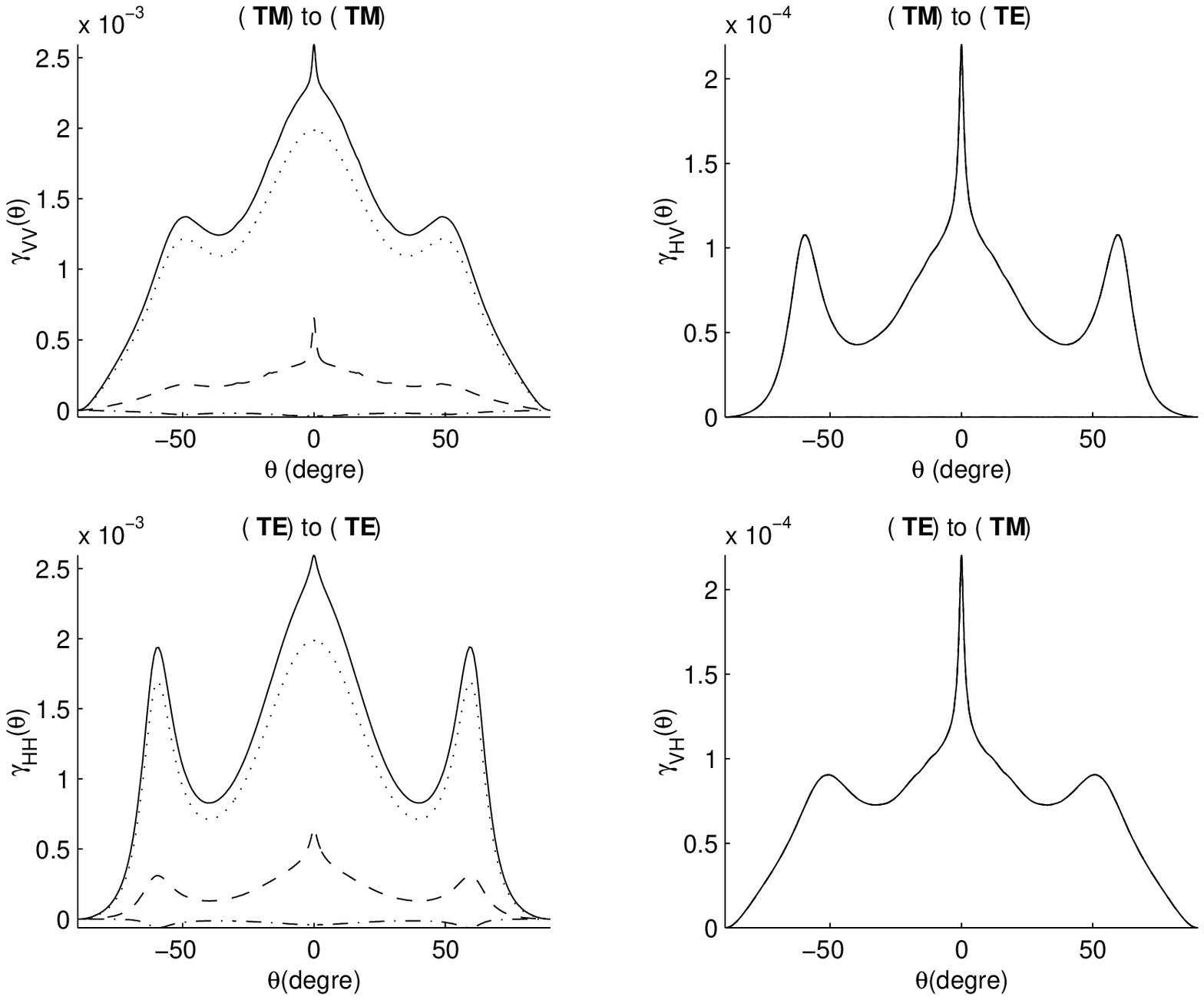}
\caption{Effect of the slab thickness, $H=1000nm$, on the configuration
shown in \Fref{SlabInf}.}
\label{SlabInf2H}
\end{figure}

The permitivities are the same as in the previous configuration,
excepted that the case $\ep_2=-\infty$ cannot be
treated with the (SPM) because the second and third order diverge.
The rms height $\sigma$ has now the value $\sigma=5\, nm$ and
$l=100\, nm$. We have not chosen $\sigma=15\, nm$ because numerically 
we have noticed that the first order term $\op{I}^{(1-1)}$ was not
greater than the second order $\op{I}^{(2-2)}$, which means that we are 
near the limit of validity of (SPM).
The perturbative development is given by \sref{Xb1}{Xb3}, and 
the guided modes are the roots of
$[X_{VV\,d}^{(0)}(p_{TM})]^{-1}$ for $(TM)$ polarization, and of
$[X_{HH\,d}^{(0)}(p_{TE})]^{-1}$ for $(TE)$  polarization.
We obtain two modes in the $(TE)$ case,  whose values are :
$p_{TE}^{1}=1.5534\,K_0$, and 
$p_{TE}^{2}=1.2727\,K_0$, the corresponding satellite peaks angles are :
$\theta_{\pm}^{12}(TE)=\pm 16.3\degre$.
For the $(TM)$ case, we have three guided modes with
$p_{TM}^{1}=1.7752\,K_0$, $p_{TM}^{2}=1.4577\,K_0$ and $p_{TM}^{3}=1.034\,K_0$,
they correspond to six possible sattelite peaks angles given by
$\theta_{\pm}^{12}(TM)=\pm 18.51\degre$, $\theta_{\pm}^{13}(TM)=\pm
47.8\degre$, and $\theta_{\pm}^{23}(TM)=\pm 25\degre$.
We see the apparition of satellite peaks only for $(TM)$ to $(TM)$ 
scattering process as shown in \Fref{SlabInf}. 
This result differs from the previous case because, on one hand,
the rough surface being not a perfect conductor we still obtain satellite peaks,
on the other hand, these satellite peaks now appear for the $(TM)$ to $(TM)$
polarization instead of $(TE)$ to $(TE)$.
This is a surprising result because 
the $(TM)$ polarization which has one more mode than the $(TE)$ one, should
decrease the amplitude of the satellite peaks for this polarization as 
it was the case with an upper rough boundary. Moreover, we see in \Fref{SlabInf2},
that the three satellite peaks can be clearly separated.
This can be explained from the fact that  there are two phenomenons which occur in
this case. The first is the same as in the previous case where the wave can excite guided modes
through the roughness which produces the enhancement of backscattering and
the satellite peaks. These effects come from the term
\begin{equation}
\alp{1}{1}\,\op{X}_{s\, \ep_1,\ep_2}^{H\,(1)}(\p{}|\p{1})\cdot\op{U}^{(0)}(\p{1})\cdot\op{V}^{10}(\p{1})
\cdot \op{X}_{s\,\ep_1,\ep_2}^{H\,(1)}(\p{1}|\p{0})
\label{exp1}
\end{equation}
in \Eref{X2d} where
$\op{U}^{(0)}(\p{1})$ have resonances for the different modes of the
guided wave. But, there is also a second phenomenon which was described
in our first example where the rough surface can excite a plasmon
mode. This appears from \Eref{X2d} with the term 
\begin{equation}
\op{X}_{s\,\ep_1,\ep_2}^{H\,(2)}(\p{}|\p{1}|\p{0})\,
\label{exp2}
\end{equation}
and subsequently in \Tref{incoh2}{ich2}. 
The localization of this mode  $\p{r}$ is given by :
\begin{equation}
[D^+_{21\,VV}(\p{r})]^{-1}=\frac{1}{\ep_2\,\alp{1}{r}+\ep_1\,\alp{2}{r}}\, .
\end{equation} 
In our case this give $||\p{r}||=1.7755\, K_0$ which is very close to the
value $p^1_{TM}=1.7752\,K_0$. So, in \Eref{ich2} the product $\odot$ 
of \eref{exp1} by \eref{exp2} can produce peaks where
$\p{1}=\pm ||\p{r}|| \ex\approx \pm p^1_{TM} \ex$ and
$(\p{0}+\p{})\cdot\ex=\pm(p^1_{TM}-p^n_{TM})$ with $n=1,2,3$.
We have effectively verified numerically that the product of this two
term can enhanced considerably the different peaks in particular the
first satellite peaks
$\theta_{\pm}^{12}$ (when $n=2$).
Now, by doubling the slab thickness, we see from \Fref{SlabInf2H} that 
the satellite peaks have disappeared due to the too many guided modes
which can be excited.
%%%%%%%%%%%%%%%%%%%%%%%%%%%%%%%%%%%%%%%%%%%%%%%%%%%%%%%%%%%%%%%%%%%%%%%%%%%%%%%
\section{Conclusions}
%%%%%%%%%%%%%%%%%%%%%%%%%%%%%%%%%%%%%%%%%%%%%%%%%%%%%%%%%%%%%%%%%%%%%%%%%%%%%%%
\label{Concl}
We have obtained four generalized reduced Rayleigh equations which 
are exact integral equations, and where one of the four unknown fields 
coming on the rough surface has been eliminated. 
These equations offer a systematic method to compute
the small perturbation development without lenghty calculations, moreover,
the scattering matrices are only two dimensional. 
All the theoretical calculations have been done up to order three 
in the height elevation which allow us to obtain all the fourth-order
cross-section terms.
We have calculated the perturbative development for three different
structures composed of a rough surface separating to semi-infinite
media, and a dielectric film where one of the two boundaries is a rough surface. 
For the first structure, the perturbative expression has
been already calculated at the third order, but our derivation offers the
advantage to be formulated in compact manner making easier numerical
computations.
For the slabs configuration we present new results. It has to be
noticed that for the case of a rough surface in the upper
position, the generalized derivation of the reduced Rayleigh equations
becomes mandatory. The numerical results show an
enhancement of the backscattering for co- and cross-polarizations in all these
cases. In the slab case, for some configurations and definite
polarizations, we have detected satellite peaks which result from
interference of different waveguide modes. This general formulation can be
extended to the configuration including two rough surfaces, and some
results will be presented in a next paper.
%%%%%%%%%%%%%%%%%%%%%%%%%%%%%%%%%%%%%%%%%%%%%%%%%%%%%%%%%%%%%%%%%%%%%%%%%%%%%%%
\begin{acknowledgments}
(AS) thanks ANRT for financial support during the preparation of
his thesis (contract CIFRE-238-98).
\end{acknowledgments}
%%%%%%%%%%%%%%%%%%%%%%%%%%%%%%%%%%%%%%%%%%%%%%%%%%%%%%%%%%%%%%%%%%%%%%%%%%%%%%%
\appendix
%%%%%%%%%%%%%%%%%%%%%%%%%%%%%%%%%%%%%%%%%%%%%%%%%%%%%%%%%%%%%%%%%%%%%%%%%%%%%%%

%%%%%%%%%%%%%%%%%%%%%%%%%%%%%%%%%%%%%%%%%%%%%%%%%%%%%%%%%%%%%%%%%%%%%%%%%%%%%%%
\section{The integration by parts}
%%%%%%%%%%%%%%%%%%%%%%%%%%%%%%%%%%%%%%%%%%%%%%%%%%%%%%%%%%%%%%%%%%%%%%%%%%%%%%%
\label{IP}
We need to calculate the following integral:
\begin{equation}\int \rmd^2 \x
\e(-i(\ku{}{1b}-\kp{}{1a})\cdot\rx)\,\nabla \hx\, .
\label{IntegPart}
\end{equation}

Since $\nabla h(x,y)$ is zero for $|x|>L/2$ or $|y|>L/2$, we can fix
the integration limits. We choose the boundary limits $x_l$ in $x$ such that
$|x_l|>L/2$, and $(\vec{u}-\vec{p})_x\, x_l=2\pi m_x$, with $m_x \in
\mathbb{Z}$. Similarly, we choose the boundary $y_l$ in $y$ such that
$|y_l|>L/2$ and $(\vec{u}-\vec{p})_y\, y_l=2\pi m_y$, with $m_y \in
\mathbb{Z}$.
Thus, the integral \eref{IntegPart} is :
\begin{align}
&\int_{-x_l}^{x_l} \int_{-y_l}^{y_l} \rmd x \rmd y \e(-i(\vec{u}-\p{}).\x)\, 
\nabla \hx\e(-i(b\alu{1}{}-a\alp{1}{})\hx) \no \\
&=\ex \int_{-y_l}^{y_l} \rmd y
\left[\frac{\e(-i(\vec{u}-\p{})\cdot\x-i(b\alu{1}{}-a\alp{1}{})\hx)}
{-i(b\alu{1}{}-a\alp{1}{})}\right]_{x=-x_l}^{x=+x_l} \no \\
&+\ey \int_{-x_l}^{x_l} \rmd
x\left[\frac{\e(-i(\vec{u}-\p{})\cdot\x-i(b\alu{1}{}-a\alp{1}{})\hx)}
{-i(b\alu{1}{}-a\alp{1}{})}\right]_{y=-y_l}^{y=+y_l} \no \\
&-\int_{-x_l}^{x_l} \int_{-y_l}^{y_l}
\frac{-\rmi(\vec{u}-\p{})}{-\rmi(b\alu{1}{}-a\alp{1}{})}
\e(-i(\vec{u}-\p{})\cdot\x-i(b\alu{1}{}-a\alp{1}{})\hx) \\
&=-\int \rmd^2\x
\frac{(\vec{u}-\p{})}{(b\alu{1}{}-a\alp{1}{})}\e(-i(\ku{}{1b}-\kp{}{1a})\cdot
\rx)\, .
\end{align}
The term in the square bracket canceled due to the choice made for
$x_l$ and $y_l$.
From the previous calculations, we can now replace $\nabla h(\x)$ by:
\begin{equation}
\nabla h(\x)\longleftrightarrow
-\frac{(\vec{u}-\p{})}{(b\alu{1}{}-a\alp{1}{})}\, .
\end{equation}
%%%%%%%%%%%%%%%%%%%%%%%%%%%%%%%%%%%%%%%%%%%%%%%%%%%%%%%%%%%%%%%%%%%%%%%%%%%%%%%
\section{Perturbative development and reciprocity condition}
%%%%%%%%%%%%%%%%%%%%%%%%%%%%%%%%%%%%%%%%%%%%%%%%%%%%%%%%%%%%%%%%%%%%%%%%%%%%%%%
\label{A2}
As was noticed by Voronovich,~\cite{Voro} the scattering operator
$\op{R}$ has a very simple law of transformation when we shift the
boundary in the horizontal direction by a vector $\vec{d}$:
\begin{equation}
\op{R}_{\x\rightarrow
h(\x-\vec{d})}(\p{}|\p{0})=\e[-i(\p{}-\p{0})\cdot\vec{d}]\,
\op{R}_{\x\rightarrow h(\x)}(\p{}|\p{0})\, ,
\label{depl}
\end{equation}
or, when we translate the surface by a vertical shift $H\,\ez$:
\begin{equation}
\op{R}_{h+H}(\p{}|\p{0})=\e[-i(\alp{0}{}+\alp{0}{0})\,H]\,\op{R}_{h}\, .
(\p{}|\p{0})
\label{translation}
\end{equation}
Now, using \eref{depl}, we can deduce some properties on the perturbative 
development of the scattering operator. The generalization of the Taylor 
expansion for a function depending on a real variable to an 
expansion depending on 
a function (which is in fact a functionnal) can be expressed in the 
following form:
\begin{equation}
\op{R}(\p{}|\p{0})=\op{R}^{(0)}(\p{}|\p{0})+\op{R}^{(1)}(\p{}|\p{0})+
\op{R}^{(2)}(\p{}|\p{0})+\op{R}^{(3)}(\p{}|\p{0})+\cdots \, ,
\end{equation}
where
\begin{align}
\op{R}^{(1)}(\p{}|\p{0})&=\intp{1}\,\op{R}^{(1)}(\p{}|\p{1}|\p{0})\,h(\p{1})\, ,
\\
\op{R}^{(2)}(\p{}|\p{0})&=\iintp{1}{2}\,\op{R}^{(2)}(\p{}|\p{1}|\p{2}|\p{0})\,
h(\p{1})\,h(\p{2})\, , \\
\op{R}^{(3)}(\p{}|\p{0})&=\iiintp{1}{2}{3}\,
\op{R}^{(3)}(\p{}|\p{1}|\p{2}|\p{3}|\p{0})\,h(\p{1})\,h(\p{2})\,h(\p{3})\, .\\
&\vdots \no
\end{align}
Applying this perturbative development on each side of \eref{depl},
and taking their functional derivative (see \citeo{Kravtsov}) defined by:
\begin{equation}
\frac{\delta^{(n)}}{\delta h(\q{1})..\delta h(\q{n})}\, .
\end{equation}
We obtain for all $n\geq 0$ in the limit $h=0$ :
\begin{equation}
\op{R}^{(n)}(\p{}|\q{1}|\cdots|\q{n}|\p{0})=\e(-\rmi(\p{}-\q{1} ....
-\q{n}-\p{0})\cdot\vec{d})\,\op{R}^{(n)}(\p{}|\q{1}|\cdots|\q{n}|\p{0})\, .
\end{equation}
We find that
\begin{equation}
\op{R}^{(n)}(\p{}|\q{1}|\cdots|\q{n}|\p{0})\propto \delta(\p{}-\q{1}
....-\q{n}-\p{0})
\end{equation}
so we can define  $\op{X}$ matrices by the relations :
\begin{align}
\op{R}^{(0)}(\p{}|\p{0})&=(2\pi)^2\,\delta(\p{}-\p{0})\,\op{X}^{(0)}(\p{0})\, ,
\\
\op{R}^{(1)}(\p{}|\p{0})&=\alp{0}{0}\,\op{X}^{(1)}(\p{}|\p{0})\,h(\p{}-\p{0})
\, , \\
\op{R}^{(2)}(\p{}|\p{0})&=\alp{0}{0}\,\intp{1}\,
\op{X}^{(2)}(\p{}|\p{1}|\p{0})\,h(\p{}-\p{1})\,h(\p{1}-\p{0})\, ,
\\
\op{R}^{(3)}(\p{}|\p{0})&=\alp{0}{0}\,\iintp{1}{2}\,
\op{X}^{(3)}(\p{}|\p{1}|\p{2}|\p{0})\,h(\p{}-\p{1})\,h(\p{1}-\p{2})\,
h(\p{2}-\p{0})\, ,\\
&\vdots \no
\end{align}
where $\alp{0}{0}$ is introduced for a matter of convenience.

Let us now make some remarks about the reciprocity condition.
If we define the anti-transpose operation by:
\begin{equation}
\begin{pmatrix}
a & b \\ 
c & d
\end{pmatrix}^{aT}=
\begin{pmatrix}
a & -c \\ 
-b & d
\end{pmatrix}\, ,
\end{equation}
the reciprocity condition for an incident and a scattered waves in the
medium 0 reads~\cite{Voro}:
\begin{equation}
{\op{R}^{aT}(\p{}|\p{0})\over \alp{0}{0}}=
{\op{R}(-\p{0}|-\p{})\over \alp{0}{}}\,.
\label{reciproque}
\end{equation}
Making use of the previous functional derivative, 
we would like to prove that each order of the perturbative development must 
satisfies this condition. It is easy to show that
\begin{equation}
\left[\op{X}^{(1)}(\p{}|\p{0})\right]^{aT}=\op{X}^{(1)}(-\p{0}|-\p{})\,,
\label{reciproq2}
\end{equation}
thus $\op{X}^{(1)}$ is reciprocal, but the same conclusion cannot be
extended to $\op{X}^{(n)}$ when $n\geq2$. 
For example, in the case $n=2$, using \eref{reciproque},
we can only deduce that:
\begin{equation}
\intp{1}\,\left[\op{X}^{(2)}(\p{}|\p{1}|\p{0})\right]^{aT}\,h(\p{}-\p{1})\,
h(\p{1}-\p{0})=\intp{1}\,\op{X}^{(2)}(-\p{0}|-\p{1}|-\p{})\,h(\p{}-\p{1})\,
h(\p{1}-\p{0})\, .
\end{equation}
From this we cannot deduce a result similar to \eref{reciproq2} for
$\op{X}^{(2)}$. This fact is well illustrated with the following
identity(which can be demonstrate with a transformation of the
integration variables)) :
\begin{equation}
\intp{1}\,(\p{}+\p{0}-2\,\p{1})\,h(\p{}-\p{1})\,h(\p{1}-\p{0})=0\, ,
\label{iden}
\end{equation}
We see that $\p{1}\longrightarrow \p{}+\p{0}-2\,\p{1}$ is not the null 
function although the integral is null.
From this we deduce that $\op{X}^{(n)}$ for $n>1$ are not unique.
Moreover in using \eref{iden} we can transform the  $\op{X}^{(n)}$
in a reciprocal form. This procedure is illustrated in the
one-dimensional case in \citeo{Mendez}, and the results for
the second-order in the electromagnetic case are given in \citeo{Voro}
and \citeo{McGurn}.

%%%%%%%%%%%%%%%%%%%%%%%%%%%%%%%%%%%%%%%%%%%%%%%%%%%%%%%%%%%%%%%%%%%%%%%%%%%%%%%

%
\listoffigures
%%%%%%%%%%%%%%%%%%%%%%%%%%%% Figures %%%%%%%%%%%%%%%%%%%%%%%%%%%%%%%%
\newpage
% figure 1
\newpage
% figure 2
\newpage
% figure 3
%\newpage
% figure 4
\begin{figure}[ht]
\input{systeme3_1.pstex_t}
\caption{A slab formed with a bottom two-dimensional rough surface
and an upper planar surface.}
\label{slabB}
\end{figure}
\newpage
% figure 5
%\newpage
\newpage
% figure 7
\begin{figure}[ht]
\input{systeme2_1.pstex_t}
\caption{A slab formed with an upper two-dimensional rough surface and
a bottom planar surface.}
\label{slabU}
\end{figure}
\newpage
\newpage
% figure 9
\newpage
% figure 10
\newpage
% figure 11
\newpage
% figure 12
\newpage
% figure 13
\newpage
% figure 14
\newpage
% figure 15
\newpage
% figure 16
\end{document}